\documentclass[aps,prx,twocolumn,showpacs,notitlepage]{revtex4-1}

\usepackage{graphicx}
\usepackage{amsmath}
\usepackage{amstext}
\usepackage{graphicx}
\usepackage{amsmath}
\usepackage{amstext}
\usepackage{amssymb}
\usepackage{amsfonts}
\usepackage{amsbsy}
\usepackage{dcolumn}
\usepackage{bm}
\usepackage{esint}
\usepackage{multirow}
\usepackage{hyperref}
\hypersetup{
    colorlinks=true,
    linkcolor=blue,
    filecolor=magenta,
    urlcolor=cyan,
}
\usepackage{cleveref}

\usepackage{mathrsfs}
\usepackage{amsfonts}
\usepackage{amsbsy}
\usepackage{dcolumn}
\usepackage{bm}
\usepackage{multirow}
\usepackage{color}
\def\ep{{\epsilon^{\mu\nu\lambda\rho} }}
  \usepackage{extarrows}

\newtheorem{defn}{Definition}

\newtheorem{conj}{Conjecture}
\usepackage{datetime}

\begin{document}
\title{Vortex-line condensation in three dimensions: A physical mechanism for bosonic topological insulators}

  \author{Peng Ye}
  \affiliation{Perimeter Institute for Theoretical Physics, Waterloo, Ontario, Canada N2L 2Y5}
  \author{Zheng-Cheng Gu}
  \affiliation{Perimeter Institute for Theoretical Physics, Waterloo, Ontario, Canada N2L 2Y5}
\begin{abstract}
%
Bosonic topological insulators (BTI) in three dimensions are symmetry-protected topological phases (SPT) protected by time-reversal and boson number conservation  {symmetries}. BTI in three dimensions were first proposed and classified by the group cohomology theory which suggests two distinct root states,  each carrying a $\mathbb{Z}_2$ index.  Soon after, surface anomalous topological orders were proposed to identify different root states of BTI, which even leads to a new BTI root state beyond the group cohomology classification.
In this paper, we  propose a universal physical mechanism via \textit{vortex-line condensation}  {from} a 3d superfluid to achieve all  {three} root states. It naturally produces bulk topological quantum field theory (TQFT) description for each root state.
Topologically ordered states on the surface are \textit{rigorously} derived by placing TQFT on an open manifold, which allows us to explicitly demonstrate the bulk-boundary correspondence. Finally, we generalize the mechanism to $Z_N$ symmetries and discuss potential SPT phases beyond the group cohomology classification.
  \end{abstract}
 \pacs{75.10.Jm, 73.43.Cd}

\maketitle

\section{Introduction}\label{sec:introduction}

 {Symmetry-protected topological states (SPT) in strongly interacting boson/spin systems have been intensely studied recently.}\cite{1DSPT,Chenlong,Chen10,Wenscience,Senthilreview}  By definition, the bulk of a SPT state only  {supports} gapped bosonic excitations but its boundary may exhibit anomalous quantum phenomena protected by global symmetry.
 As such, the usual three-dimensional fermionic topological insulators (FTI) \cite{TI1,TI2,TI3,TI4,TI5,TI6,TIexp} can be literally viewed as a fermionic SPT state.\cite{explainspt}  Haldane spin chain, which was proposed decades ago, is a typical example of SPT states in 1d.\cite{Haldane1,Haldane2,Haldane3,Haldane4,1DSPT,Pollmann10}
 {Mathematically, given} both spatial dimension and symmetry group $G$ as input data, one can apply the ``group cohomology theory with $\mathbb{R}/\mathbb{Z}$ coefficient'' to systematically classify   SPT states.\cite{Chenlong}      Another mathematical tool ``cobordism'' has also been applied and some nontrivial  SPT states beyond group cohomology theory have been proposed  {recently}.\cite{cobordism,footnotewencoho}
In addition to the above classification frameworks, a surge of broad interest has been shown from different approaches.\cite{2d1,2d2,2d3,2d4,2d5,2d6,2d61,2d7,2d8,2d9,2d10,2d11,wengauge,2d12,2d13,2d14,2d15,2d16,2d17,2d18,2d19,2d20,WGW,2d14,2d61,2d19,3d1,3d2,3d3,3d4,3d5,3d6,WGW,3d7,3d8,Geraedts,alicea}


 In this paper, we will study bulk topological quantum field theory (TQFT) of ``bosonic topological insulators'' via the so-called ``vortex-line condensation'' mechanism. {As a bosonic analog of the well-known FTIs,  the so-called ``bosonic topological insulators'' (BTI) were proposed first based on the group cohomology theory\cite{Chenlong}}.
  By definition, a BTI state is a nontrivial SPT state protected by U(1)$\rtimes$Z$^T_2$ symmetry in three dimensions. Here, U(1) symmetry denotes the conservation of boson number, while, time-reversal symmetry Z$^T_2$ acts on bosons as $\mathcal{T}^2=1$ in the bulk. In the framework of group cohomology theory, such SPT states are classified by $\mathbb{Z}_2\times \mathbb{Z}_2$\cite{Chenlong,2d14,footnoteroots}. In other words, in comparison with the single $\mathbb{Z}_2$ index in FTIs of free fermions \cite{TI1,TI2,TI3,TI4,TI5,TI6,TIexp} which corresponds to the even / odd number of Dirac cones on the surface, there are two independent $\mathbb{Z}_2$ indices to label distinct BTI states and each index allows us to define a so-called ``BTI root state''.\cite{footnoteroots} It was nicely proposed in \cite{3d3} that surface of BTI supports topological order that: (i) respects symmetry and (ii) cannot be realized on a 2d plane alone unless symmetry is broken. In the following, let us briefly review all BTI root states.

The nontrivial phenomena of  the \textit{first BTI root state} can be characterized by its surface Z$_2$ topological order where both $e$ and $m$ quasiparticles carry half-charge.
In addition, if Z$^T_2$ is explicitly broken on such a surface and the bulk is fabricated in a slab geometry, one may expect a nontrivial electromagnetic response   featured by quantum Hall effect with odd-quantized Hall conductance on the surface and bulk Witten effect with $\Theta=2\pi\,\text{mod}\,4\pi$\cite{3d3,3d5,witten1,witten2,witten3}, which is different from  $\Theta=\pi\,\text{mod}\,2\pi$ in FTI states of free fermions.\cite{qhz}  BTI labeled by this $\mathbb{Z}_2$ index has been studied in details via fermionic projective construction and dyon condensation.\cite{3d5}
The  physical signature of the \textit{second BTI root state}  is characterized by its surface Z$_2$ topological order where both $e$ and $m$ quasiparticles are Kramers' doublets.
Surprisingly, it has been recently known that there is a new $\mathbb{Z}_2$ index that is beyond group cohomology classification.\cite{3d3,3d4,cobordism,footnotewencoho}  As the
\textit{third BTI root state}, it supports a nontrivial surface with the so-called ``all-fermion'' Z$_2$ topological order  where all three nontrivial quasiparticles are self-fermionic and mutual-semionic.
Remarkably, an exactly solvable lattice model for this BTI has been proposed via the so-called Walker-Wang approach,\cite{3d1,ww12} which   confirms the existence of the third $\mathbb{Z}_2$ index.


\begin{table*}
 \begin{tabular}[t]{cccc}
\hline\hline~\\
 \begin{minipage}[t]{1.5in}\textbf{BTI}\end{minipage} &\begin{minipage}[t]{3in} \textbf{Bulk TQFT}  \end{minipage}& \begin{minipage}[t]{2.3in}\textbf{Surface Topological Order} \end{minipage}  \\\\\hline \\
   \begin{minipage}[t]{1.2in} The first BTI root state in Sec. \ref{sec:pbf2}\end{minipage}&\begin{minipage}[t]{3in}
    $\frac{K^{IJ}}{4\pi}b^I_{\mu\nu}\partial_\lambda a^J_\rho \ep$ \end{minipage} &\begin{minipage}[t]{2.3in} {Z$_p$ ($p=\text{even}$)}   topological order with exotic electric charge assignment.  U(1)  is defined in an unusual way.\end{minipage} \\~\\
    \begin{minipage}[t]{1.2in}The second  BTI root state in Sec. \ref{sec:pbf1}\end{minipage}&\begin{minipage}[t]{3in}
     $\frac{K^{IJ}}{4\pi}b^I_{\mu\nu}\partial_\lambda a^J_\rho \ep$ \end{minipage} &\begin{minipage}[t]{2.3in} Z$_2$   topological order where both $e$ and $m$ carry Kramers' doublets.   Z$^T_2$ is defined in an unusual way.\end{minipage} \\
~\\
     \begin{minipage}[t]{1.2in}The third  BTI root state in Sec. \ref{sec:btifti} {  (beyond cohomology)}\end{minipage}&\begin{minipage}[t]{3in}
      $\frac{K^{IJ}}{4\pi}b^I_{\mu\nu}\partial_\lambda a^J_\rho \ep+\frac{\Lambda^{IJ}}{16\pi}b_{\mu\nu}^Ib_{\lambda\rho}^J\ep$\end{minipage} &\begin{minipage}[t]{2.3in} Z$_2$   topological order where all $e$, $m$, and $\epsilon$ quasiparticles are fermionic. U(1)$\rtimes $Z$^T_2$ is defined in a usual way.\\~ \end{minipage}
 ~\\
      \hline\hline
  \end{tabular}
  \caption{A brief summary of main results in Sections \ref{sec:btifti}-\ref{sec:pbf1}. As the 1d chiral Luttinger liquid theory which can be derived by putting Chern-Simons action on a 2d disk with a 1d boundary, surface of each BTI root state is also rigorously derived by putting the bulk field theory on an open 3d manifold with a 2d boundary. Symmetry transformations (both U(1) and time-reversal Z$^T_2$) are  rigorously defined in the bulk. 
  The first and second BTI root states are within group cohomology classification and obtained by changing definition of either U(1) symmetry or Z$^T_2$ symmetry in the bulk. The third BTI root state is beyond group cohomology and its realization requires an addition of ``cosmological constant term'' $b\wedge b$ term. More details (e.g. the integer matrices $K^{IJ}$, $\Lambda^{IJ}$) are present in the main text.}\label{table2}
  \end{table*}

\section{Overview}
{Despite of much progress  in diagnosing surface phenomena of BTI states, throughout the paper, we stress that a well-defined bulk theory and bulk definition of symmetry  are very crucial towards a  {controllable} understanding of the surface quantum states.} This concern is  also highlighted in the conclusion section of \cite{witten3}. If the bulk theory is unknown, the uniqueness of a proposed surface state is generically  unclear. More concretely, one may  understand the importance of a bulk definition through the following two aspects. Firstly, given a 2d  state that cannot be symmetrically realized in any 2d lattice model, it does not necessarily mean that the state can be realized on the surface of a 3d SPT phase.
Secondly, when a surface phase transition occurs, the bulk doesn't necessarily experience a bulk phase transition, implying that a many-to-one correspondence between boundary and bulk is generically possible. Incidentally, a many-to-one correspondence was   studied in quantum Hall states with high Landau levels.\cite{cano14,nayak13} In the present paper, we will also show that the first BTI root state (Sec. \ref{sec:pbf2})  exhibits a many-to-one correspondence. It generalizes the aforementioned descriptions of surface topological order where only Z$_2$ topological order is allowed.

\begin{figure}[t]
\centering
\includegraphics[width=8.5cm]{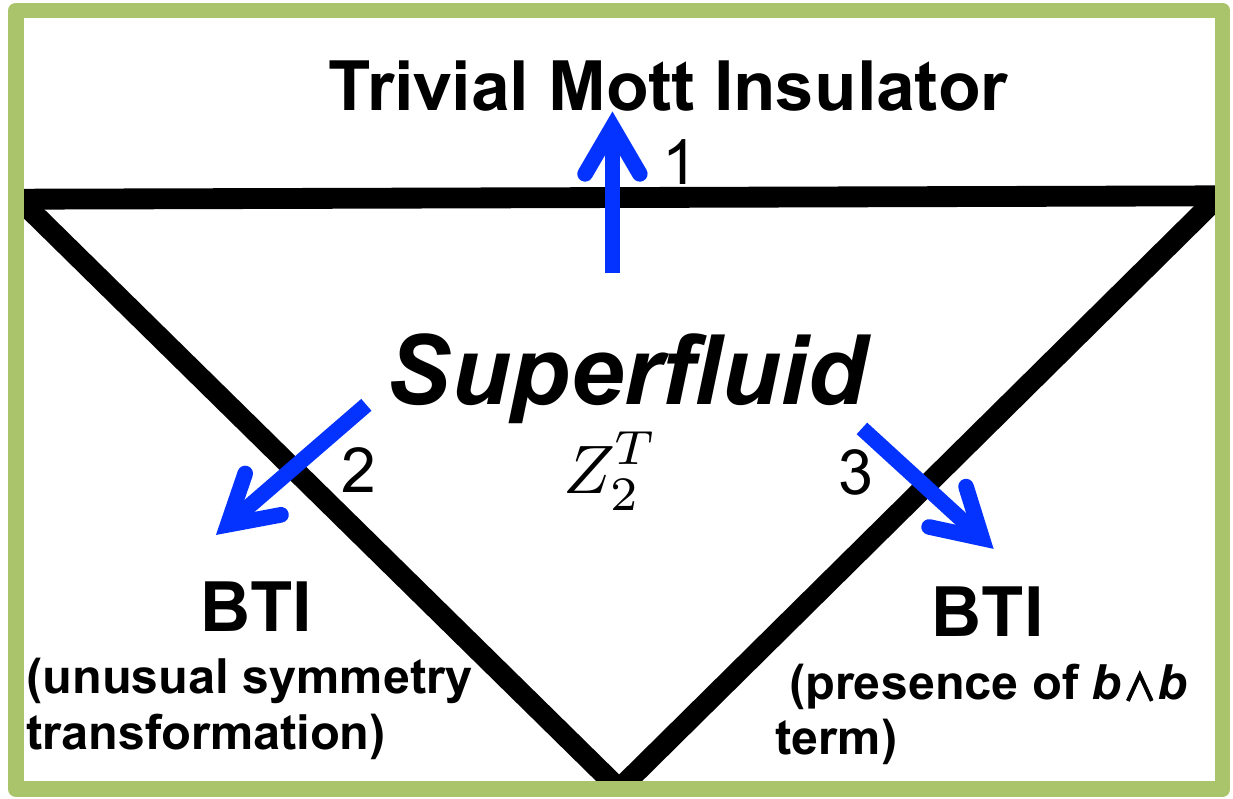}
\caption{Phases obtained by vortex-line condensation.  In the \textit{Phase transition-1}, U(1) symmetry (i.e. boson number conservation) is restored from superfluid to a trivial Mott insulator  by condensing strings (\textit{i.e.} 2$\pi$-vortex-lines). Thus, the trivial Mott insulator phase is formed by vortex-line condensation with $b\wedge \mathrm{d}a$ type bulk field theory description.
In the \textit{Phase transition-2}, strings are also condensed and the bulk field theory is also $b\wedge \mathrm{d}a$ type (see   Sec. \ref{sec:pbf2}, \ref{sec:pbf1}). But the resultant Mott phase is a nontrivial SPT state (i.e. bosononic topological insulators, BTI) since    either U(1) or Z$^T_2$ symmetry transformations is defined in an unusual way. Thus, we end up with two different BTI root states. In the \textit{Phase transition-3}, strings are condensed in the presence of a nontrivial linking Berry phase term, or more precisely, an nontrivial multicomponent $b\wedge b$ type term. The nontrivial Mott phase is a  BTI phase obtained in Sec. \ref{sec:btifti}, which is a SPT root state beyond group cohomology classification and supports ``all-fermion'' Z$_2$ surface topological order.  Here, Z$^T_2$ denotes time-reversal symmetry with $\mathcal{T}^2=1$. }
\label{figure_triangle}
\end{figure}
One way to derive bulk   field theory is the so-called ``hydrodynamical approach'' that we will apply in this paper.   Let us   introduce this approach by briefly reviewing its application in  fractional quantum Hall effect (FQHE).
 FQHE is a strongly correlated many-body quantum system, and it is technically hard to derive the bulk low-energy field theory  by directly performing renormalization group analysis.  However, one may apply the hydrodynamical approach whose  main principle is to study collective modes of Hall system at low energies.  Since the bulk is gapped, it is sufficient to take quantum fluctuations of density $\rho$ and current $\vec{j}$ into account. 
Along this line of thinking, the bulk  Chern-Simons theory is obtained, which  encodes various  ``topological data'' of the incompressible Hall state,  such as modular $\mathcal{S}_m$ and $\mathcal{T}_m$ matrices, chiral central charge $c^-$, and other properties of edge conformal field theory (CFT).\cite{Wenbook,km1,km2,km3,km4,km5,km6,km7}   
Very recently, TQFT protected by global symmetry has also been intensively studied. For example, Lu and Vishwanath \cite{2d8}  imposed   global symmetry to Chern-Simons theory and successfully classified some 2d SPT states protected by Abelian symmetry group.  {The success of such a hydrodynamical approach in 2d SPT\cite{2d8}    motivates us to  develop a ``universal hydrodynamical approach'' for SPT phases in 3d.

It is intricate to tackle bulk TQFT of all SPT states in 3d, which is far beyond the scope of the present work. In this paper,  we restrict our attention to investigating  the dynamic topological quantum field theory of aforementioned BTI states through considering  exotic ``vortex-line condensations'', which is pictorially illustrated in Fig. \ref{figure_triangle}. Here vortex-lines mean the configuration of topological line defects  in 3d superfluid states, e.g. helium-4. Such a vortex-line condensate state is shown to be described by a topological action in the form of
 \begin{equation}
S_{\rm{top}}=i\frac{K^{IJ}}{2\pi}\int b^I\wedge \mathrm{d}a^J+i\frac{\Lambda^{IJ}}{4\pi}\int b^I\wedge b^J, \label{action}
\end{equation}
where $a^I$ are usual 1-form U(1) gauge fields and $b^I$ are 2-form   U(1) gauge fields.\cite{wedge_convention} $K$ and $\Lambda$ are some $N\times N$ integer matrices that will be elaborated in details in main texts, $I,J=1,2,\cdots N$.
Surprisingly, we find  {that} such a simple physical picture is sufficient to produce all three root states of 3d BTI.
More concretely, we find that the first two  BTI  root states within group cohomology classification can be achieved  through a pure $b\wedge \mathrm{d}a$ type term where symmetry transformations (either  Z$^T_2$ or U(1)) are defined in an unusual way. The third BTI root state beyond group cohomology classification requires the presence of a nontrivial $b\wedge b$   term whose physical meaning is illustrated in Fig. \ref{figure_sheet}. A brief summary is given by Table \ref{table2}. {Our physical approach to 3d BTIs avoids the complication of advanced mathematical topics like} cohomology theory and cobordism theory. And we believe that this physical picture will shed light on a more challenging question in the future: how to design microscopic interaction terms that can realize those proposed BTI states.

 The vortex-line condensation picture is also formally generalized to some other symmetry groups, e.g., unitary Z$_N$ group as discussed in Sec. \ref{sec:znspt}.  It turns out that there are more potentially nontrivial SPT states beyond the group cohomology classification in 3d. The bulk dynamical quantum field theory of the new Z$_N$ SPT state is described by a single-component version of Eq. (\ref{action}) with non-vanishing $\Lambda$.  Based on our results, we  conjecture that:
 \begin{conj}
 All SPT phases in 3d described by a  $b\wedge b$ term are beyond the group cohomology classification.
\end{conj}

The remainder of the paper is organized as follows.  Sec. \ref{sec:origin} is devoted to understand the microscopic origins of bulk dynamical TQFT for 3d BTI through the hydrodynamical approach. In this section, we will start with a superfluid state in 3d and derive the TQFT description of the vortex-line condensate.  In Sec. \ref{sec:general}, some useful  properties of the TQFT are studied when symmetry is not taken into account.  By taking symmetry into account, the BTI state beyond group cohomology classification\cite{footnotewencoho} is obtained in Sec. \ref{sec:btifti} where a nontrivial presence of $b\wedge b$ term plays an important role. In Sec. \ref{sec:pbf2}, \ref{sec:pbf1}, the remaining two BTI root states within group cohomology classification are obtained via a pure $b\wedge \mathrm{d}a$ type term by defining either   Z$^T_2$ or  U(1) symmetry transformation in an unusual way. In Sec. \ref{sec:znspt}, we show that $b\wedge b$ term might also lead to   Z$_N$ SPT phases beyond group cohomology class. A concluding remark is made in Sec. \ref{sec:con}. Some future directions are also proposed.

\section{Hydrodynamical approach to topological quantum field theory}\label{sec:origin}
\subsection{3d superfluid state and its dual description}\label{sec:sfdual}

The exotic states to be discussed in this paper are built up from a well known parent state: ``3d superfluid (SF) state''   described by:
\begin{align}
\mathcal{L}=\frac{\rho}{2}(\partial_\mu \theta)^2\,\label{xymodel}
\end{align}
at low energies.
Here, $\rho$ is the superfluid density. $\theta$ is the U(1) phase angle of the superfluid. The spatial gradient of $\theta$ costs energy such that a spatially uniform value of $\theta$ is picked up in the ground state, rendering a \textit{spontaneous} symmetry breaking of global U(1) symmetry group (\textit{i.e.} the particle number conservation of bosons).  In order to capture the periodicity of $\theta$ in the continuum field theory,  we may express $\theta$ in terms of smooth part and singular part: $\theta=\theta^{\rm s}+\theta^{\rm v}$\,.
By substituting this $\theta$ decomposition into  Eq. (\ref{xymodel}) and introducing a Hubbard-Stratonovich auxiliary vector field $ \mathcal{J}^\mu$\cite{duality1,duality2,duality3}, one may express Eq. (\ref{xymodel}) in the following equivalent formalism:
\begin{align}
\mathcal{L}=\frac{1}{2\rho}( \mathcal{J}_\mu)^2+ i\mathcal{J}^\mu(\partial_\mu \theta^{\rm s}+\partial_\mu\theta^{\rm v})\,\label{eqj}
\end{align}
which goes back to  Eq. (\ref{xymodel}) once $\mathcal{J}^\mu$ is integrated out.
 It is obvious that $ \mathcal{J}^\mu$ can be interpreted as supercurrent of the 3d SF state.
Integrating out $\theta^{\rm s}$ leads to a constraint $\delta(\partial_\mu \mathcal{J}^\mu)$ in the path integral measure.
%
This constraint can be resolved by introducing a 2-form non-compact U(1) gauge field $b_{\mu\nu}$: $\mathcal{J}^\mu\xlongequal{\text{def.}}\frac{1}{4\pi}\ep \partial_\nu b_{\lambda\rho}\,$.
Both the physical quantity $\mathcal{J}^\mu$ and the Lagrangian (\ref{eqj}) are  invariant under  the usual smooth gauge transformation:
\begin{align}
b_{\mu\nu}\rightarrow b_{\mu\nu}+\partial_{[\mu}\xi_{\nu]}\,,\nonumber
 \end{align}
 where $\xi_\mu$ is a smooth 4-vector. ``$\partial_{[\mu}\xi_{\nu]}$'' stands for ``$\partial_{\mu}\xi_{\nu}-\partial_{\nu}\xi_{\mu}$''.
Eventually, the Lagrangian (\ref{eqj}) is transformed to the following  gauge theory:
\begin{align}
\mathcal{L}=\frac{1}{48\pi^2\rho}{h}^{\mu\nu\lambda}h_{\mu\nu\lambda}+\frac{i}{2}b_{\mu\nu}\Sigma^{\mu\nu}\,,\label{sfdual}
\end{align}
where the field strength $h_{\mu\nu\lambda}$ is a rank-3 antisymmetric tensor: $h_{\mu\nu\lambda}\xlongequal{\text{def.}}\partial_{\mu}b_{\nu\lambda}+\partial_{\nu}b_{\lambda\mu}+\partial_{\lambda}b_{\mu\nu}
$. In order to simplify notation, ``$\mathcal{L}_h$'' is introduced via
\begin{align}
\mathcal{L}_h\xlongequal{\text{def.}}\frac{1}{48\pi^2\rho}{h}^{\mu\nu\lambda}h_{\mu\nu\lambda}\,\label{bmaxwell}
\end{align}
which is the Maxwell term of the two-form U(1) gauge field $b_{\mu\nu}$.
The vortex-line (\textit{i.e.} string) current operator $\Sigma_{\mu\nu}$ which is antisymmetric is defined through the singular $\theta^{\rm v}$:
\begin{align}
\Sigma^{\mu\nu}\xlongequal{\text{def.}}\frac{1}{2\pi}\ep \partial_\lambda \partial_\rho \theta^{\rm v}\,\label{sfdual1}
\end{align}
which is generically nonzero for nontrivial homotopy mapping.
The gauge transformation shown above automatically ensures that there is a continuity equation for $\Sigma_{\mu\nu}$, i.e. $\partial_\nu \Sigma^{\mu\nu}=0$.
Hereafter, the nouns ``strings'', ``vortex-lines'', and ``closed loops'' are used interchangeably. The vortex-line configuration is very dilute in superfluid. The factor $\frac{1}{2}$ in the coupling term $\frac{1}{2}b_{\mu\nu}\Sigma^{\mu\nu}$ naturally arises as a standard convention for the antisymmetric tensor field coupling in 3+1d space-time.

\subsection{Trivial Mott insulators realized by condensing vortex-lines (strings)}\label{sec:trivialstring}
Considering strong correlation effects (like Hubbard interactions), we expect that passing through a  critical point where the tension of vortex-lines decreases to zero, the string configuration (denoted by the path integral measure $\mathscr{D}\Sigma$) will be proliferated energetically. In other words,  ``vortex-line condensation''\cite{MS,Franz07} sets in.
 The path-integral formalism of vortex-line condensation was ever given in Refs. \cite{Franz07,Rey89,stringbook}. Here, we shall not go into technical details but  briefly review the basic method. A single string can be described by repametraization-invariant Nambu-Goto action. A wave function $\Psi$ can be introduced in quantum theory of strings. Similar to the usual quantum theory of particles, after promoting the quantum mechanics of single-string to field theory of many-strings, $\Psi$ will be viewed as the creation operator (in operator formalism) or the quantum amplitude (in path-integral formalism) of a given string configuration.

In condensate of bosons, the ground state is formed by equal-weight superposition of all kinds of boson configurations in real space, which leads to a macroscopic wave-function.  The amplitude fluctuation of is gapped but the phase fluctuation $\theta$ is gapless and governed by Eq.(\ref{xymodel}). Likewise, once vortex-line condensation sets in, all vortex-line configurations have the same quantum amplitude $\Psi$. In contrast to condensate of bosons,  the U(1) phase of $\Psi$ of vortex-line condensate is given by a Wilson line $e^{i\int dx_\mu \Theta^\mu}$ and governed by the Lagrangian given below:
\begin{align}
\mathcal{L}=&\frac{1}{2}\phi_0^2 \left( \partial_{[\mu}\Theta_{\nu]}- b_{\mu\nu}\right)^2+\mathcal{L}_h \,,\label{mini}
\end{align}
where the antisymmetrization symbol is defined as usual: $\partial_{[\mu}\Theta_{\nu]}\xlongequal{\text{def.}}\partial_{\mu}\Theta_{\nu}-\partial_{\nu}\Theta_{\mu}$. $|\phi_0|^2$ is the ``phase stiffness'' of the vortex-line condensate. The presence of dynamical gauge field $b_{\mu\nu}$ gaps out the gapless phase fluctuation from $\Theta_\mu$.
One may split the phase vector into smooth part $\Theta^s_\mu$ and singular part $\Theta^{\rm v}_\mu$:
$
\Theta_\mu=\Theta^{\rm s}_\mu+\Theta^{\rm v}_\mu\,\, $, where, $\int d^3\mathbf{r}\frac{1}{4\pi}\nabla\cdot\nabla\times \mathbf{\Theta}^{\rm v}\in\mathbb{Z}\,.$ Therefore, the gauge group of $b_{\mu\nu}$ is compactified by absorbing $ {\Theta}_\mu^{\rm v}$.   Note that, in the dual Lagrangian (\ref{sfdual}) in SF phase, $b_{\mu\nu}$ is not compact.

Based on Eq. (\ref{mini}), we may formally perform duality transformation in this vortex-line condensation to obtain a  $b\wedge da $-term where $a$ is 1-form gauge field. For this purpose, let us introduce a Hubbard-Stratonovich auxiliary tensor field $\Sigma_{\mu\nu}$ :
\begin{align}
\mathcal{L}=i\frac{1}{2}\Sigma_{\mu\nu}(\partial^{[\mu} \Theta^{\nu]} -b^{\mu\nu})+\mathcal{L}_h +\frac{1}{8\phi_0^2}\Sigma_{\mu\nu}\Sigma^{\mu\nu}\,,\label{dualL2}
\end{align}
where the physical interpretation of $\Sigma_{\mu\nu}$ is same as the one defined in Eqs. (\ref{sfdual},\ref{sfdual1}).

Integrating over $\Theta^{\rm s}_\mu$ in Eq. (\ref{dualL2}) yields a constraint $\delta(\partial_\nu \Sigma^{\mu\nu})$ in the path integral measure.  This constraint   can be resolved by introducing a 1-form non-compact U(1) gauge field $a_\mu$:
\begin{align}
\Sigma^{\mu\nu}\xlongequal{\text{def.}}-\frac{1}{2\pi}\ep \partial_\lambda a_\rho\label{Sigma1}
\end{align}
indicating that $a_\mu$ field strength is physically identified as the ``supercurrent'' $\Sigma_{\mu\nu}$ of vortex-lines.
The vector field $a_\mu$ is a gauge field since under the gauge transformation:
\begin{align}
a_\mu\rightarrow a_\mu+\partial_\mu\eta\,,\label{equation:gaugea}
\end{align}
the physical observable $\Sigma^{\mu\nu}$ is invariant.  $a_\mu$ takes values smoothly on whole real axis so that $a_\mu$ is non-compact, thereby, leading to $\partial_\nu \Sigma^{\mu\nu}=0$.
 Then, the dual formalism of Lagrangian (\ref{mini}) is given by:
\begin{align}
\mathcal{L}=&i\frac{1}{4\pi}a_\mu \ep \partial_\nu {b}_{\lambda\rho}+i a_\mu j_v^\mu+\mathcal{L}_h+\frac{1}{64\pi^2\phi_0^2}f_{\mu\nu}f^{\mu\nu}\,,\label{ablambda}
\end{align}
 The field strength tensor $f_{\mu\nu}\xlongequal{\text{def.}} \partial_{[\mu} a_{\nu]}$ as usual. The `monopole' current of the string condensate is given by $j_v^\mu=-\frac{1}{4\pi}\partial_\nu\partial_{[\lambda}\Theta^v_{\rho]}\ep$. We may redefine $b$ by absorbing $\text{d}\Theta^v$: $b_{\mu\nu}-\partial_{[\mu}\Theta^v_{\nu]}\rightarrow b_{\mu\nu}$.
 Once removing the irrelevant Maxwell terms $\mathcal{L}_h$ and $f_{\mu\nu}f^{\mu\nu}$ at low energies, we end up with the following topological $BF$ Lagrangian:
\begin{align}
\mathcal{L}=i\frac{1}{4\pi}\ep b_{\mu\nu}\partial_\lambda a_\rho\label{trivialMIL}
\end{align}
 As expected, the coefficient $\frac{1}{4\pi}$ of the first term in Eq. (\ref{trivialMIL}) indicates that there is no ground state degeneracy (GSD)\cite{Wtop,WNtop,Wrig,H8483,ASW8422,H8285,Wedgerev} on a 3-torus $\mathbb{T^3}$.\cite{bf1,bf2,bf3} In this sense, the bulk state has no intrinsic topological order\cite{Wtop,WNtop,Wrig,H8483,ASW8422,H8285,Wedgerev}. In terms of exterior products, the term can be rewritten as $\frac{1}{2\pi}b\wedge \mathrm{d}a$ where $\mathrm{d}a\equiv f$ is a 2-form field strength tensor.
%

\subsection{Adding a vortex-line (string) linking Berry phase term into the trivial Mott insulator}
In the following, we attempt to explore the possibility of nontrivial Mott insulators.  To begin, we add a topological Berry phase term  into Eq. (\ref{mini}) to describe a potential nontrivial ``topological vortex-line condensate'':
\begin{align}
\mathcal{L}=&\frac{1}{2}\phi_0^2\left( \partial_{[\mu}\Theta^s_{\nu]}- b_{\mu\nu}\right)^2+\mathcal{L}_h   \nonumber\\
&-i\frac{\Lambda}{16\pi}\ep \left( \partial_{[\mu}\Theta^s_{\nu]}-b_{\mu\nu}\right)\left( \partial_{[\lambda}\Theta^s_{\rho]}-b_{\lambda\rho}\right) ,\label{miniplus}
\end{align}
where,  $b$ is redefined by absorbing $\text{d}\Theta^v$: $b_{\mu\nu}-\partial_{[\mu}\Theta^v_{\nu]}\rightarrow b_{\mu\nu}$.
By applying the duality transformation, this topological vortex-line condensate can be equivalently described by the following `$BF+BB$' TQFT:
\begin{align}
\mathcal{L}_{\rm{top}}=&i\frac{1}{4\pi}\ep b_{\mu\nu}\partial_\lambda a_\rho+i\frac{\Lambda}{16\pi}\ep b_{\mu\nu} b_{\lambda\rho}\,,\label{eq:LBH}
\end{align}
In details,  in Eq. (\ref{miniplus}), the term $\sim d\Theta^{\rm s}\wedge d\Theta^{\rm s}$ is a total derivative term and can be neglected. Then, by introducing a Hubbard-Stratonovich auxiliary tensor field $\Xi^{\mu\nu}$ (antisymmetric), Eq. (\ref{miniplus}) is transformed to:
 \begin{align}
\mathcal{L}=&i\frac{1}{2}\Xi^{\mu\nu} (\partial_{[\mu}\Theta^{\rm s}_{\nu]}- b^{\mu\nu}) +\frac{1}{8\phi^2_0}\Xi_{\mu\nu}\Xi^{\mu\nu}+i\frac{\Lambda}{8\pi}\ep  \partial_{[\mu}\Theta^{\rm s}_{\nu]}b_{\lambda\rho}\nonumber\\
&  -i\frac{\Lambda}{16\pi}\ep b_{\mu\nu}b_{\lambda\rho}+  \mathcal{L}_h \,\nonumber\\
 =&i\Theta^{\rm s}_{\mu}\partial_\nu\left(\Xi^{\mu\nu} +\frac{\Lambda}{4\pi}\ep b_{\lambda\rho} \right)- i\frac{1}{2}\Xi_{\mu\nu}b^{\mu\nu}\nonumber\\
 & -i\frac{\Lambda}{16\pi}\ep b_{\mu\nu}b_{\lambda\rho} +\frac{1}{8\phi^2_0}\Xi_{\mu\nu}\Xi^{\mu\nu}+\mathcal{L}_h\,.\label{lag2}
\end{align}
Integrating out $\Theta_\mu^{\rm s}$ leads to the conservation constraint:
$\partial_\nu\left(\Xi^{\mu\nu} +\frac{\Lambda}{4\pi}\ep b_{\lambda\rho} \right)=0
$
which can be resolved by introducing a 1-form non-compact U(1) gauge field $a_\mu$:
\begin{align}
\Xi^{\mu\nu}\xlongequal{\text{def.}}-\frac{1}{2\pi}\ep \partial_\lambda a_\rho-\frac{\Lambda}{4\pi}\ep b_{\lambda\rho}\,.\nonumber
\end{align}
This is a `modified' version of Eq. (\ref{Sigma1}) where $b\wedge b$-term is absent.    Plugging this expression into the second term in Eq. (\ref{lag2}) yields the   topologically invariant Lagrangian Eq. (\ref{eq:LBH}).

In Eq. (\ref{eq:LBH}), only topological terms are preserved. $\mathcal{L}_h$, which is defined in Eq. (\ref{bmaxwell}),  is the Maxwell kinetic term of $b_{\mu\nu}$ with scaling dimension  more irrelevant than the two topological terms in Eq. (\ref{eq:LBH}).
 In addition, we consider the phase region that is deep in the string condensation phase and far away from the phase boundary between SF and string condensate. As such, the ``phase stiffness'' $|\phi_0|^2\rightarrow \infty$ is taken.
 The first term in Eq. (\ref{eq:LBH}) is the standard $b\wedge \mathrm{d}a$ term that already exists in Sec. \ref{sec:trivialstring}. What is new here is the  second term  ``$b\wedge b$''. It  was previously introduced  in mathematical physics.\cite{horowitz89}  It is also applied  to loop quantum gravity\cite{smolin} with cosmological constant. Its physical meaning is pictorially shown in  Fig. \ref{figure_sheet}.
 \begin{figure*}[t]
\centering
\includegraphics[width=13cm]{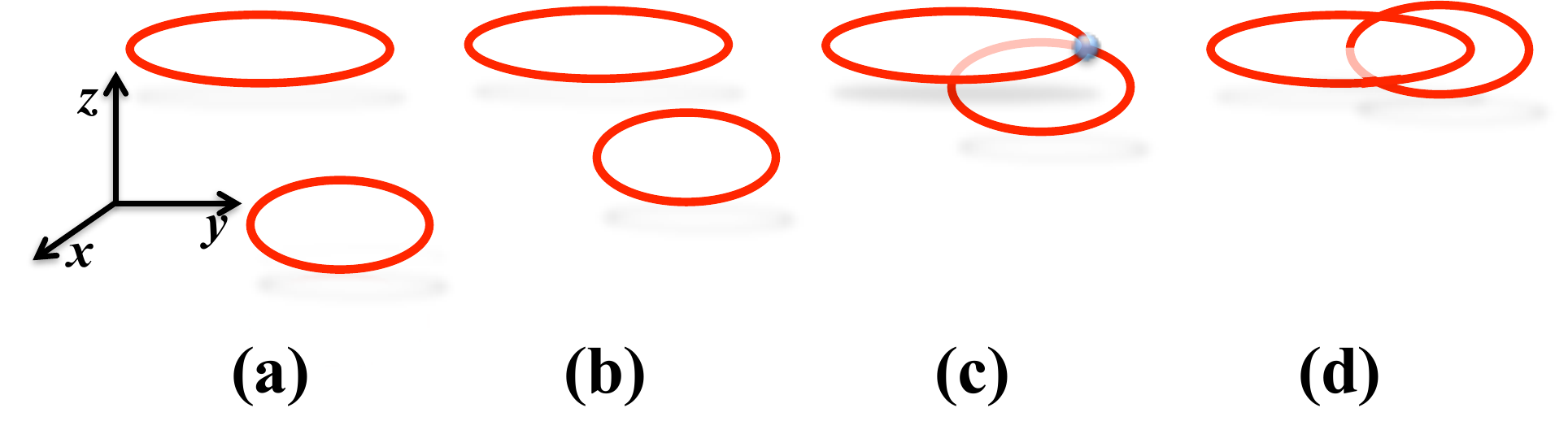}
\caption{Physical meaning of $b\wedge b$ topological term. The larger loop denotes a vortex-line that is static  and located on $xy$-plane. The smaller loop is perpendicular to $xy$-plane, parallel to $yz$-plane, and moves toward $z$-direction. There are four snap-shots shown in this figure from left to right. The blue dot in the third snap-shot denotes the intersection of two loops. In the fourth snap-shot, two loops are eventually linked to each other. $b\wedge b$ term will contribute a phase at the third snap-shot of the unlinking-linking process.}
\label{figure_sheet}
\end{figure*}
The above topologically invariant Lagrangian is gauge invariant under the following gauge transformations:
\begin{align}
b_{\mu\nu}\longrightarrow b_{\mu\nu}+\partial_{[\mu}\xi_{\nu]}\,\,,\,\,a_{\mu}\longrightarrow a_{\mu}+\partial_\mu \eta-\Lambda\xi_\mu\,.\label{gaugetransform0}
\end{align}

Formally, the above single-component theory can be generalized into a multi-component theory:
\begin{align}
\mathcal{L}_{\rm{top}}= i\frac{K^{IJ}}{4\pi}\ep b^{I}_{\mu\nu}\partial_\lambda a^J_{\rho}+ i\frac{\Lambda^{IJ}}{16\pi}\ep b^I_{\mu\nu}b^J_{\lambda\rho}
\,.
\label{Lagrangian}
\end{align}
In terms of exterior products, the action  (\ref{action}) is obtained.
Without loss of generality, it is sufficient to consider symmetric matrix $\Lambda^{IJ}$ and assume   $K^{IJ}$  to be an identity matrix of rank-$N$, \textit{i.e.}
\begin{align}
K=\text{diag}(1,1,\cdots,1)_{N\times N}=\mathbb{I}\,\label{eq:K}
\end{align}
with $I,J=1,2,\cdots, N$.
$\{a^I_{\mu}\}$ are non-compact 1-form U(1) gauge fields and $\{b^I_{\mu\nu}\}$ are compact 2-form U(1) gauge fields, respectively, as a straightforward generalization of the above one-component theory. The above topologically invariant Eq.(\ref{Lagrangian}) Lagrangian with the particular $K$ matrix Eq.(\ref{eq:K}) is the central result of this paper (its abstract form in terms of exterior products is given by Eq. (\ref{action})), and we will use it to describe all   BTI phases as well as some new Z$_N$ SPT phases. Physically, such a multi-component theory can be viewed as  a collection of many 3d trivial Mott insulators mutually entangled via  $\Lambda$ term. Hereafter, we will also call indices $I,J,\cdots$
\textit{flavor indices}.

%
%
%

\section{General properties of the topological quantum field theory}\label{sec:general}
In this section, we will study the multi-component topological quantum field theory defined by Eq. (\ref{Lagrangian}) in a general setting without any global symmetry implementation.
 {In this section as well as Sec. \ref{sec:btifti}, all analyses are done by implicitly assuming $K$ takes the form in Eq (\ref{eq:K}) unless otherwise stated.}
( e.g. the gauge transformation and   $\mathbb{GL}(N,\mathbb{Z})$ transformation in Sec. \ref{sec:gaugetransform} are valid for general $K$.)

 \subsection{Gauge transformation and Bulk $\mathbb{GL}(N,\mathbb{Z})$ transformation}\label{sec:gaugetransform}
 As aforementioned, the presence of $b\wedge b$-term drastically changes the gauge structures.
 The gauge transformation of the multi-component theory Eq. (\ref{Lagrangian}) is given by:
 \begin{align}
 b^{I}_{\mu\nu}\longrightarrow b^I_{\mu\nu}+\partial_{[\mu}\xi^I_{\nu]}\,\,,\,\,a^I_{\mu}\longrightarrow a^I_{\mu}+\partial_\mu \eta^I-( K^{-1}\Lambda)^{II'}\xi^{I'}_\mu\,\label{gaugetransform}
 \end{align}
which  generalizes  Eq. (\ref{gaugetransform0}).

To see the gauge transformation more clearly, we may reexpress the first two topological terms (denoted by $\mathcal{L}_{\rm top}$) in Lagrangian (\ref{Lagrangian}) as:
 \begin{align}
 \mathcal{L}_{\rm top}=&i\frac{\Lambda^{IJ}}{32\pi}\ep\left(b^I_{\mu\nu}+(\Lambda^{-1}K)^{II'}\partial_{[\mu}a^{I'}_{\nu]}\right)\nonumber\\
&~ ~~ ~~\cdot~\left(b^J_{\lambda\rho}+(\Lambda^{-1}K)^{JJ'}\partial_{[\lambda}a^{J'}_{\rho]}\right)\,.\label{eq:bbgauge}
 \end{align}
 From this expression, one may examine the correctness of Eq. (\ref{gaugetransform}) easily.
To obtain this equivalent expression, we have applied the following two facts: (\textit{i}) closed space-time manifold is taken. (\textit{ii}) $a_\mu$ is non-compact such that the term $\sim d a \wedge d a$ is a total derivative.

%
%

 In analog to 2+1d Chern-Simons theory, we can   perform two independent general linear ($\mathbb{GL}$) transformations represented by matrices $W, M\in\mathbb{GL}(N,\mathbb{Z})$, on gauge fields $b_{\mu\nu}^I$ field and $a_{\mu}^I$, respectively. $\mathbb{GL}$ transformations keep the quantization of gauge charges of both gauge fields unaffected:
  \begin{align}
  \underline{ b}^I_{\mu\nu}= (W^{-1})^{IJ} b^J_{\mu\nu}\,\,,\underline{a}^I_{\mu} =(M^{-1})^{IJ}a^J_{\mu}\,,\label{eq:gltrans}
 \end{align}
 where $W, M$ are two $N$ by $N$ matrices with integer-valued entries and $|\text{det}W|=|\text{det}M|=1$.
  These transformations are nothing but a ``relabeling'' of the same low energy physics.  After the transformations, a new set of parameters  $(\underline{K},\underline{\Lambda}$ can be introduced via:
 \begin{align}
& \underline{K}=W^TKM\,,  \underline{\Lambda}=W^T\Lambda W\, \label{eq:wm}
 \end{align}
 which leads to a new Lagrangian in the same form as Eq. (\ref{Lagrangian}).
 $W$ and $M$ are two independent $\mathbb{GL}$ transformations. In any basis, $|\rm{det}K|$ rather than $\rm{det} K$ is invariant. Therefore, our choice Eq. (\ref{eq:K}) is universal once bulk topological order is absent.

\subsection{Quantization conditions on path-integral field variables}\label{sec:periodicity}
 For simplicity, let us  merely consider one-component theory. It is straightforward to generalize all of results obtained below to multi-component theory by adding component indices.
The action is given by:
 \begin{align}
 S=i\frac{1}{4\pi}\int d^4x b_{\mu\nu}\partial_\lambda a_\rho \ep+i \frac{\Lambda}{16\pi}\int d^4x b_{\mu\nu}b_{\lambda\rho}\ep\,. \label{equation:onecomponenttheory}
 \end{align}
 It is known that the \textit{classical} action $S$ alone is not enough to define a quantum system. One must properly define the partition function $\mathsf{Z}$ where the path-integral measure should be properly defined in addition to the classical action. First, a formal integration over the one-form gauge field $a_\mu$ will lead to a flat connection constraint on the two-form gauge field with local flatness $\mathrm{d}b=0$. Second, we note that there is a shift in the microscopic justification of the single-component action based on the mechanism `vortex-line condensation': $b_{\mu\nu}+\partial_{[\mu}\Theta^{\rm v}_{\nu]}\rightarrow b_{\mu\nu}$ which leads to a quantization condition of $b_{\mu\nu}$.
 Quantitatively,  we have the following   {quantization condition} on $b$  on a closed 2d manifold $\mathcal{S}$ embedded in 4d spacetime:
 \begin{align}
 \varoiint_\mathcal{S} b=\varoiint_\mathcal{S} \text{d}\Theta^{\rm v}=\int\!\!\!\!\!\int \!\!\!\!\! \int_\mathcal{V} \text{d}\, \text{d} \Theta^{\rm v}=2\pi\times \,\text{integer}\,,\label{equation:higgsed}
 \end{align}
 where, $\mathcal{S}=\partial\mathcal{V}$.
In deriving the third BTI root state (see Table \ref{table2}) that needs a nontrivial   $b\wedge b$,  we will use this   {quantization condition}   which is obtained based on the microscopic origin ``vortex-line condensation''.  As a matter of fact, once the topological terms $BF+BB$ are formally derived from our microscopic origin, we may extend our discussion on the topological field theory to a more general background. In other words, this   {quantization condition} on $b$ field can  be derived in a more mathematically rigorous way by formally introducing compactness of both gauge fields which are fundamentally constrained by two conditions (ie. Eqs. \ref{condition1_b} and \ref{condition1_a}) given in Appendix \ref{appendix_gsd}.  In Sec. \ref{sec:pbf2} and Sec. \ref{sec:pbf1} where the derivation of the first and second BTI root states (see Table \ref{table2}) are present respectively, we will continue to use this general method.


By using the condition (\ref{equation:higgsed}), we may derive the following periodic shift: (more details are present in Appendix \ref{derivingequation:quantizationzn})
 \begin{align}
 \Lambda\rightarrow \Lambda+1\,.\label{equation:u1periodicity}
 \end{align}
 Thus, $\Lambda$ is compactified to a finite region $[0,1)$. Further, one may consider large gauge transformation which leads to quantization of $\Lambda\in\mathbb{Z}$. The derivation of this result may be consider as a special case of Eq. (\ref{equation:quantizationzn}) with $N=1$. Therefore, without time-reversal symmetry,  any allowed nonzero $\Lambda$ is identified as $\Lambda=0$.

 More generally, the above periodicity shift of the single-component theory is also applicable  to the diagonal entries $\Lambda^{II}$ of a generic multi-component theory. For off-diagonal entries, the results are still unchanged. For example, since $\Lambda^{13}$ term and $\Lambda^{31}$ term are equal to each other ($\Lambda^{13}=\Lambda^{31}$), the total actions of the mixture of $b^1_{\mu\nu}$ and $b_{\mu\nu}^3$ are actually: $2\times \frac{\Lambda^{13}}{16\pi}\int d^4x \ep b^1_{\mu\nu}b^3_{\lambda\rho}$. We note that there are eight equivalent copies in Eq. (\ref{eq:deriveshift}) where the two $b$ fields are the same. However, in $\Lambda^{13}$ term where the two $b$ fields are different, there are only four equivalent copies. Overall, loss and gain are balanced such that the periodicity of $\Lambda^{13}$ is still 1.
%

In addition,  the condition (\ref{equation:higgsed}) can also be applied to derive the surface theory.  Due to  this condition, $b^I$ is locally flat and a new 1-form  {compact} gauge field  $\tilde{a}^I_{\mu}$ for each index $I$ can be introduced via:
\begin{align}
    b^I_{\mu\nu}\xlongequal{\text{def.}}\partial_{[\mu} \tilde{a}^I_{\nu]}=\tilde{f}^I_{\mu\nu}\,, \label{resolution_b}
    \end{align}
where $\tilde{f}^{I}_{\mu\nu}$ is the field strength tensor of $\tilde{a}^I_{\mu}$.
  For the sake of convenience,  $\mu,\nu,\lambda=0,1,2$ is implicitly assumed in all of surface variables. The compactness of $\tilde{a}^I$ can be easily understood by substituting it into the  {quantization condition} of $b^I$. Then, the magnetic flux of $\tilde{a}^I$ piercing a closed $\mathcal{S}$ is allowed to be nonzero which means that monopoles of $\tilde{a}^I$ are allowed so that $\tilde{a}^I$ is compactified. After substituting this expression into $b\wedge b$-term in Eq. (\ref{Lagrangian}), we end up with the following surface Lagrangian:
 \begin{align}
\mathcal{L}_{\partial}=\frac{\Lambda^{IJ}}{4\pi}\epsilon^{\mu\nu\lambda}\tilde{a}^I_\mu \partial_\nu \tilde{a}^J_\lambda\,.\label{LambdaCS}
\end{align}

\section{Bosonic topological insulators in the presence of $b\wedge b$ topological term }\label{sec:btifti}
In this section, we will implement symmetry and consider $b\wedge b$ term and end up with the third BTI root state (the last row in Table \ref{table2}) that is beyond group cohomology classification.

\subsection{Definition of time-reversal  transformation }\label{sec:mbs}
Given the microscopic origin of the topological quantum field theory Eq. (\ref{Lagrangian}) in Sec. \ref{sec:origin}, we know that $b_{\mu\nu}$ is minimally coupled to vortex-lines (\textit{i.e.} strings), while, $a_{\mu}$ is minimally coupled to bosonic particles. As a result, the time-reversal transformations of all gauge fields and excitations can be consistently defined in the following usual way (the spatial directions are denoted by $i=1,2,3$):
\begin{align}
&\mathcal{T}{a}^I_0\mathcal{T}^{-1}= a^I_0\,\,,\mathcal{T}{a}^I_i\mathcal{T}^{-1}= -a^I_i\,\,,\nonumber \\
&\mathcal{T}{j}_0\mathcal{T}^{-1}=j_0\,\,, \mathcal{T}{j}_i\mathcal{T}^{-1}= -j_i\,\,,\label{timerule0}\\
&\mathcal{T}{b}^{I}_{0,i}\mathcal{T}^{-1}=-b^{I}_{0,i}\,\,,\mathcal{T}{b}^{I}_{i,j}\mathcal{T}^{-1}= b^{I}_{i,j}\,\,,\nonumber\\
&\mathcal{T}{\Sigma}_{0,i}\mathcal{T}^{-1}=-\Sigma_{0,i}\,\,,\mathcal{T}{\Sigma}_{i,j}\mathcal{T}^{-1}= \Sigma_{i,j}\,,\label{timerule}
\end{align}
where every flavor transforms in the same way.

The pure $b \wedge f$-term (\textit{i.e.} $b\wedge b$ term is absent)  in Eq. (\ref{Lagrangian}) is invariant under these transformation rules.  The definition of time-reversal transformation in Eqs. (\ref{timerule0}) and (\ref{timerule})  implies that a pure $b\wedge \mathrm{d}a$ term  \textit{necessarily} leads to a trivial SPT state for the reason that bulk topological order is absent and symmetry transformation is defined in a usual way.
The possibility of nontrivial SPT states arising from pure $b\wedge \mathrm{d}a$ term  will be  discussed in Sec. \ref{sec:pbf2}, \ref{sec:pbf1}  where either U(1) or  time-reversal symmetry has to be modified unusually, as shown in Fig. \ref{figure_triangle}.  Therefore, in the present Sec. \ref{sec:btifti}, $\Lambda=0$ always results in a trivial SPT state  and nontrivial SPT state requires a nontrivial $\Lambda$ matrix.

 \subsection{Quantization condition implemented by time-reversal symmetry}\label{section_half}
  Under the time-reversal symmetry transformation (\ref{timerule}),  $b\wedge b$-term is transformed to:
$-\frac{\Lambda^{IJ}}{16\pi}\ep b^I_{\mu\nu}b^J_{\lambda\rho}\,.
$ Superficially, this sign change implies that the ground state of topological field theory labeled by $\Lambda$ always ``breaks'' time-reversal symmetry.  However, we will show that the periodicity shift  of $\Lambda^{IJ}$    provides a chance for restoring time-reversal symmetry.   We have derived a periodic shift (\ref{equation:u1periodicity})   on $\Lambda$ where merely U(1) symmetry is considered. A new problem arises:  \textit{Is    (\ref{equation:u1periodicity})   still valid in the presence of time-reversal symmetry?}
%

Let us reconsider a simple one-component theory shown in Eq. (\ref{equation:onecomponenttheory}). In the presence of time-reversal symmetry,     the space-time manifold becomes un-oriented such that topological response can be probed.\cite{cobordism,Wencoho}.  {A simple understanding is that normal vector at each spacetime point changes sign under time-reversal symmetry such that a time-reversal invariant system requires that the each spacetime point is directionless.}  In un-oriented space-time manifold, a $\pi$ cubic flux of 2-form gauge field $b_{\mu\nu}$ becomes the minimally allowed value.  To have a physical picture for the flux quantization condition on un-oriented manifold, we can consider the simplest case -- a flux insertion process for a Mobius strip. Very different from a cylinder, where the inserted flux must be in unit of $2\pi$, the Mobius strip allows the inserted flux to be in unit of $\pi$, namely, $\oint_{L} A_\mu dl_\mu=\pi\times \text{integer}$ because a particle on a Mobius strip must pick up an even winding number to travel back to its origin.
In this sense, if we  still use the same notations as  {in Appendix \ref{sub1}}, $\mathcal{N}_{0x}$ and $\mathcal{N}_{yz}$ now are half-integers instead of integers. The last line in Eq. (\ref{eq:deriveshift}) now equals to $\frac{1}{2}\Lambda \pi\times\text{integer}$ such that the periodicity of $\Lambda$ now is enhanced to:
\begin{align}
\Lambda\rightarrow \Lambda+4\,.\label{equation:periodicitytr}
\end{align}
 Let us move on to the off-diagonal entires, e.g., $\Lambda^{13}$ in a multi-component theory. At present, there are $N$ components of ``topological vortex-line condensations'' which, superficially, implies that there are $N$ U(1) charge conservation symmetries. However, in our physical system, only one U(1) should be taken into account.  Then, when we evaluate the sum of $\Lambda^{13}$ and $\Lambda^{31}$ terms,  either $b^1$ gauge group or $b^2$ forms a $\pi$ cubic flux, not both. Therefore, the periodicity of $\Lambda^{13}$ is enhanced from 1 to 2: i.e. $\Lambda^{13}\rightarrow \Lambda^{13}+2$.
In summary, all the above results indicate that  $\Lambda^{IJ}$ in the presence of time-reversal symmetry take the following values:
\begin{align}
\Lambda^{II}=0,\pm 2\,,\Lambda^{IJ}=0,\pm 1 \text{  (for $I\neq J$)}\,.\label{equation:terminators}
\end{align}
This quantization condition is protected by time-reversal symmetry.

 We note that SPT states \textit{(including both trivial and nontrivial states)} are defined by the following two common conditions: (i) bulk has no intrinsic topological order; (ii) bulk state respects symmetry.
 The condition-(i) is always satisfied in our construction since GSD=1 as shown in Sec. \ref{sec:gaugetransform} where  $K=\mathbb{I}$. If $\Lambda$ entries are defined under the requirement of  Eq. (\ref{equation:terminators}), the condition-(ii) is also satisfied. Thus, there are infinite number of $\Lambda$ matrices that satisfy Eq. (\ref{equation:terminators}) and can be viewed as  SPT states with U(1)$\rtimes$Z$^T_2$ symmetry.
    But which are trivial and which are nontrivial? For example, is $\Lambda=2$ a trivial or nontrivial SPT? In subsequent discussions, we will aim to answer this question.

 \subsection{Trivial SPT states with $|\text{det}\Lambda|=1$}\label{sec:t}
All $\Lambda$ matrices ($K=\mathbb{I}$ is implicit all the time) that satisfy the quantization conditions (\ref{equation:terminators}) are SPT states. Generically, with an open boundary condition, the surface phenomena of SPT states are expected to capture information of triviality and non-triviality.  Therefore, one may wonder what are nontrivial signatures of surface phenomena?  Conceptually, one should first find the set of physical observables that describe the surface physics:
\begin{defn}
\textbf{Surface physical observables}. \\ The physical observables of Eq. (\ref{LambdaCS}) are composed by: ground state degeneracy, self-statistics and mutual statistics of gapped quasiparticles. All these information can be read out from   modular $\mathcal{S}_m$ and $\mathcal{T}_m$ matrices. Since there are no further 1d boundary, chiral central charge $c^-$ is not an observable on the surface. Notice that, for a topological phase defined on a 2d plane, the physical observables are $\mathcal{S}_m$, $\mathcal{T}_m$, and $c^-$.
\label{remarkobs}
\end{defn}

Now our question is changed to: how can we use these physical observables to tell a nontrivial SPT from a trivial SPT? The essential physics is the so-called ``obstruction'' or ``anomaly''. More precisely, we define it as:
\begin{defn} \textbf{Obstruction (Quantum Anomaly)}. \\
By ``{obstruction}'', we mean that the set of physical observables (defined in Definition \ref{remarkobs}) of the surface theory cannot be reproduced on a 2d plane by any local bosonic lattice model with symmetry.  Otherwise, the obstruction is free. Nontriviality of a SPT state corresponds to the presence of  obstruction.
\label{definition2dplane}
\end{defn}
 Now, we are at the position to distinguish trivial and nontrivial SPT states. We first consider a subset of $\Lambda$ matrices that satisfy Eq. (\ref{equation:terminators}):
\begin{align}
|\text{det}\Lambda|=1\,.\label{eq:detunit}
\end{align}
Mathematically, any $\Lambda$ matrix in this subset can be expressed in terms of two ``fundamental blocks'', namely,
 $\Lambda_{\rm t1}$, and, $\Lambda_{\rm t2}$ given by Cartan matrix of E$_8$ group:\cite{cartan,nayak13,wiki,kitaevannal}
            \be
            \Lambda_{\rm t1}=\begin{pmatrix}
      0 & 1\\
        1 & 0
            \end{pmatrix}  \,,~~
 \Lambda_{\rm t2}=   \left(\begin{matrix}
      2 & 1 & 0 & 0 & 0 & 0 & 0 & 0\\
      1 & 2 & 1 & 0 & 0 & 0 & 0 & 0 \\
      0 & 1 & 2 & 1 & 0 & 0 & 0 & 1 \\
      0 & 0 & 1 & 2 & 1 & 0 & 0 & 0 \\
      0 & 0 & 0 & 1 & 2 & 1 & 0 & 0 \\
      0 & 0 & 0 & 0 & 1 & 2 & 1 & 0 \\
      0 & 0 & 0 & 0 & 0 & 1 & 2 & 0 \\
      0 & 0 & 1 & 0 & 0 & 0 & 0 & 2 \\
    \end{matrix}\right) \,.\nonumber
\ee
The subscript ``$\rm{t}$'' in ``$\Lambda_{\rm t1}$'' and ``$\Lambda_{\rm t2}$'' stands for ``trivial'' (to be explained below). One can   show that all  $\Lambda$ matrices that satisfy Eq. (\ref{equation:terminators}) and (\ref{eq:detunit}) can be expressed as the following ``canonical form'', namely, a direct sum of several $\Lambda_{\rm t1}$ and $\pm \Lambda_{\rm t2}$ up to an arbitrary $\mathbb{GL}$ transformation:
\begin{align}
\Lambda=W^T(\Lambda_{\rm t1}\oplus\Lambda_{\rm t1}\oplus\cdots\oplus\pm\Lambda_{\rm t2}\oplus\pm\Lambda_{\rm t2}\oplus\cdots)W\,.\label{trivialspt1}
\end{align}
Since $\mathbb{GL}$ transformation $W$ doesn't affect physical observables, whether SPT states in this subset are trivial or not essentially depends on the properties of the two fundamental blocks.

  $\Lambda_{\rm t1}$ gives a trivial SPT state (i.e. a trivial Mott insulator) since its surface physical observables are    trivial gapped boson excitations and nothing else. Such surface state can be realized on a 2d lattice model with symmetry. According to Definition \ref{definition2dplane}, the obstruction is free and thus the 3d bulk state is trivial.

In contrast to $\Lambda_{\rm t1}$ whose chiral central charge $c_-=0$, the very feature of $\Lambda_{\rm t2}$ is that it has an ``irreducible'' value of $c_-=8$ which plays a role of the ``generator'' of all trivial $\Lambda$ matrices that admit nonzero $c_-$.  We call $\pm \Lambda_{\rm t2}$ ``$c$-generators''. By ``irreducible'', we mean that one can prove that  $c_-=8$ is the minimal absolute value of all  $\Lambda$ matrices that satisfy Eq. (\ref{equation:terminators}) and (\ref{eq:detunit}). Since $c_-\neq0$, one may wonder if $\Lambda_{\rm t2}$ surface state breaks time-reversal symmetry once it is laid on a 2d plane alone. To solve this puzzle, we should, again, focus attentions to \textit{physical observables} on the surface rather than the formal Lagrangian in Eq. (\ref{LambdaCS}).  On the grounds that a surface is a 2d closed manifold by definition, there is no further 1d edge so that  $c_-$ is not detectable on the surface, which is also summarized in Definition \ref{remarkobs}. Thus, $\Lambda_{\rm t2}$  still gives a trivial SPT state.

Technically, the   triviality of   $\Lambda_{\rm t2}$ can also be understood by using the following Gauss-Milgram sum formula\cite{cano14,GM1,GM2}
\begin{align}
\frac{1}{\sqrt{|\text{det}\Lambda|}}\sum_{a}e^{i2\pi J_a}=e^{i2\pi ic^{-}/8}\label{eqGM}
\end{align}
for bosonic Abelian topological   phase defined on a 2d plane. Here, $a$ denotes quasparticles and $J_a$ is topological spin which is determined by the modular $\mathcal{T}_m$ matrix. The diagonal entries $T^{aa}_m=e^{i2\pi J^a}$ in quasiparticle basis.   The chiral central charge $c^{-}$ is determined only modulo 8. Let us consider $\Lambda_{\rm t2}$ surface state. According to  Definition \ref{remarkobs}, the surface physical observables are determined by $\mathcal{S}_m$ and $\mathcal{T}_m$. The two modular matrices are given by: $\mathcal{S}_m=1\,,\,\mathcal{T}_m=1\,
$, both of which are just a number. Thus, the surface is nothing but a trivial gapped bosonic phase with time-reversal symmetry and supporting only trivial identity particles.
Then,   one may wonder if the time-reversal symmetry can be preserved  when we define such a set of physical observables on a 2d plane because on a 2d plane we need to further consider chiral central charge which is a signature of time-reversal breaking. The answer is Yes. The reason is that all possible  states on a 2d plane with the same $\mathcal{S}_m$ and same $\mathcal{T}_m$ have $c^-=0$ mod $8$ due to Eq. (\ref{eqGM}). Since $c^-=0$ is a solution, a time-reversal symmetric state on a 2d plane is achievable without any difficulty. Thus, according to Definition \ref{definition2dplane}, $\Lambda_{\rm t2}$ labels a trivial SPT state.

  \subsection{Trivial SPT states with $|\text{det}\Lambda|>1$}\label{sec:t1}

 The above discussion leads to a set of trivial SPT states defined by Eq. (\ref{equation:terminators}) and Eq. (\ref{eq:detunit}). All of these trivial states do not admit  topological order on the surface. How about topologically ordered surface (\textit{i.e.} $|\text{det}\Lambda|>1$)?

While the quantization conditions (\ref{equation:terminators}) guarantee that the bulk is symmetric, surface might break symmetry. If symmetry is manifestly broken on the surface, such a surface state can be realized on a 2d plane, which is free of obstruction from symmetry requirement. The corresponding bulk state is a trivial SPT state.
Therefore, hereafter we will merely focus on the symmetry preserving surface state. But a question arises: how can we judge  the symmetry is preserved on the surface?

A symmetric surface governed by Eq. (\ref{LambdaCS}) with $\Lambda$ matrix  must describe the same set of physical observables as those  on the surface with $-\Lambda$-matrix. In order to leave the surface physical observables unaffected under time-reversal (\textit{i.e.} $\Lambda\rightarrow-\Lambda$),  one obvious way is to consider the following equivalence under $\mathbb{GL}$ transformation, namely, ``$\mathbb{GL}$-equivalence'':
\begin{align}
W^T\Lambda W=-\Lambda, \,\, \exists \,W\in \mathbb{GL} (N,\mathbb{Z})\,\label{eq:lambdagl}
\end{align}
which does not change physical observables in Definition \ref{remarkobs}. We introduce the following symbol
\begin{align}
\Lambda \xlongequal{\mathbb{GL}}-\Lambda\label{eq:symbol}
\end{align} to denote this equivalence relation. However, such a topological order has no obstruction if it is defined on a 2d plane with symmetry since such a $\mathbb{GL}$ transformation can also be regularly performed on a 2d plane. Thus, the only way toward nontrivial SPT states is to find a new method such that (i) it can transform $\Lambda$ to $-\Lambda$ leaving all physical observables unaffected, and, (ii) it is forbidden to be regularly done on a 2d plane. This is what we shall do in Sec. \ref{sec:extended} where an ``extended $\mathbb{GL}$ transformation''  Eq. (\ref{eq:eGL0}) is defined.

Before proceeding further, let us give some examples. The simplest one is $\Lambda=2$ which can   be neither connected to $\Lambda=-2$ via Eq. (\ref{eq:lambdagl}) nor Eq. (\ref{eq:eGL0}), so that the corresponding bulk state is a trivial SPT with symmetry-breaking surface. Another example is $\Lambda=\left(\begin{smallmatrix}
      2 & 0\\
        0 & -2
            \end{smallmatrix} \right)$ which can be connected to $\Lambda=\left(\begin{smallmatrix}
      -2 & 0\\
        0 & 2
            \end{smallmatrix} \right)$ via Eq. (\ref{eq:lambdagl}), so that the corresponding bulk state is a trivial SPT with symmetry-preserving surface.

\subsection{Extended $\mathbb{GL}$ transformation and nontrivial SPT states}\label{sec:extended}

In order to  obtain nontrivial SPT states, again, we resort to the fact that the chiral central charge on the surface is not a physical observable as discussed in Sec. \ref{sec:t}. We may relax the $\mathbb{GL}$ transformation by arbitrarily adding fundamental blocks like $\Lambda_{\rm t1}$ and $\pm\Lambda_{\rm t2}$ defined in Sec. \ref{sec:t} along the diagonal entries of $\Lambda$ matrix. This stacking is legitimate since all fundamental blocks correspond to trivial SPT states which do not induce phase transitions.

Technically, we perform a so-called ``extended $\mathbb{GL}$ transformation'' via the following equivalence relation:
\begin{align}
&W^T(\Lambda \oplus\Lambda_{\rm t1}\oplus\Lambda_{\rm t1}\cdots)W=(-\Lambda)\oplus \pm\Lambda_{\rm t2}\oplus \pm \Lambda_{\rm t2}\cdots, \,\, \nonumber\\
&\exists \,W\in \mathbb{GL} (N',\mathbb{Z})\,\label{eq:eGL0}
\end{align}
which also leaves physical observables in Definition \ref{remarkobs} unaffected.
Here, the left-hand-side contains $\frac{(N'-N)}{2}$ $\Lambda_{\rm t1}$ matrices, while, the right-hand-side contains $\frac{(N'-N)}{8}$ ``$\pm\Lambda_{\rm t2}$'' matrices. Here, $\Lambda$ is   $N\times N$ as usual. The extended $\mathbb{GL}$ transformation for $\Lambda$ means that adding several fundamental blocks both to $\Lambda$ and $-\Lambda$ results in two new matrices, we can then connect these two new matrices, \textit{i.e.} $(\Lambda \oplus\Lambda_{\rm t1}\oplus\Lambda_{\rm t1}\cdots)$ and $(-\Lambda)\oplus \pm\Lambda_{\rm t2}\oplus \pm \Lambda_{\rm t2}\cdots$, by performing a $\mathbb{GL}$ transformation $W$.

 Since  $ \left(\Lambda_{\rm t2}\oplus( -\Lambda_{\rm t2})\right)\xlongequal{\mathbb{GL}}\sum^8\oplus\Lambda_{\rm t1} $, the $\pm$ signs reduce to an overall sign, \textit{i.e.}
\begin{align}
&W^T(\Lambda \oplus\Lambda_{\rm t1}\oplus\Lambda_{\rm t1}\cdots)W=(-\Lambda)\oplus \left[  \pm\left(\Lambda_{\rm t2}\oplus  \Lambda_{\rm t2}\cdots\right)\right], \,\, \nonumber\\
&\exists \,W\in \mathbb{GL} (N',\mathbb{Z})\,.\label{eq:eGL}
\end{align}
If $\Lambda$ and $-\Lambda$ are connected to each other via an extended $\mathbb{GL}$ transformation, we define the symbol:
\begin{align}
\Lambda \xlongequal{\mathbb{eGL}}-\Lambda\nonumber
\end{align}
to denote their equivalence relation.

We find that there is a \textit{unique solution} that supports nontrivial SPT. It is the Cartan matrix of SO(8) group, denoted by $\Lambda_{\rm so8}$:\cite{cartan}
\be
   \Lambda_{\rm so8}=  \left( \begin{matrix}
      2 & 1 & 1 & 1\\
        1 & 2 & 0& 0\\
          1 & 0 & 2 & 0\\
            1 & 0 & 0 & 2\\
    \end{matrix} \right)\,.\label{eq:lambda8matrix}
\ee
 More precisely, the following transformation exists:
 \begin{align}
 &W^T(\Lambda_{\rm so8} \sum^4 \oplus\Lambda_{\rm t1})W=(-\Lambda_{\rm so8})\oplus  \Lambda_{\rm t2}, \,\, \nonumber\\
&\exists \,W\in \mathbb{GL} (12,\mathbb{Z}) \,.\nonumber
\end{align}
Instead of directly looking for the explicit matrix form of $W$, the existence of $W$ can be   justified by checking the equivalence of the physical observables (see Definition \ref{remarkobs}) between $(\Lambda_{\rm so8} \sum^4 \oplus\Lambda_{\rm t1})$-surface-Chern-Simons theory and $(-\Lambda_{\rm so8})\oplus  \Lambda_{\rm t2}$-surface-Chern-Simons theory. Indeed, both share the same excitation spectrum that are formed by four distinct gapped quasiparticles,  and, they share the same real $\mathcal{S}_m$ and $\mathcal{T}_m$ matrices:
\begin{align}
\mathcal{S}_m=  \left( \begin{matrix}
       1& 1 & 1 & 1\\
        1 & 1 & -1& -1\\
          1 & -1 &1 & -1\\
            1 & -1 & -1 & 1\\
    \end{matrix}  \right)\,,\mathcal{T}_m=  \left( \begin{matrix}
       1& 0 & 0 & 0\\
        0& -1 & 0& 0\\
          0 & 0 & -1 & 0\\
            0 & 0 & 0 & -1\\
    \end{matrix}  \right)\,.
 \end{align}
From the above $\mathcal{S}_m$ and $\mathcal{T}_m$, we may read much information. In addition to the trivial boson excitations (identity quasiparticles), there are three distinct fermions denoted by $f_1,f_2,f_3$. Braiding $f_i$ around $f_j$ ($\forall i,j$, $i\neq j$) once leads to an Aharonov-Bohm phase $e^{i\pi}=-1$.  Consequently, through the equivalence relation Eq.(\ref{eq:eGL0}), we end up with an SPT state whose surface has topological order.

 Furthermore,  it is   a nontrivial SPT, \textit{i.e.} {bosonic topological insulator (BTI)}.  To see its non-triviality clearly, one may attempt to look for obstruction defined in Definition \ref{definition2dplane}.  We put the set of physical observables (given by the modular $\mathcal{S}_m$ and $\mathcal{T}_m$ matrices of $\Lambda_{\rm so8}$) on a 2d plane. According to the Gauss-Milgram sum formula Eq. (\ref{eqGM}), the chiral central charge $c^-=4$ mod $8$. Therefore, all states on a 2d plane neccesarilly have nonzero chiral central charge, indicating that time-reversal symmetry is necessarily broken. Thus, such an obstruction defined in Definition \ref{definition2dplane} gives a 3d nontrivial SPT state  labeled by $\Lambda_{\rm so8}$.


In summary, we derive a BTI state from our bulk field theory where a nontrivial multi-component $b\wedge b$ term plays an essential role. We stress that this BTI state is obtained and examined rigorously from bulk to boundary step-by-step.


   \section{Bosonic topological insulators from pure $b\wedge \mathrm{d}a$ term$-$(I)}\label{sec:pbf2}
  In the above discussions, we consider the time-reversal transformation defined by Eqs. (\ref{timerule0}) and (\ref{timerule}) such that the pure $b\wedge \mathrm{d}a$ term always describes  trivial SPT states, i.e.  ``trivial Mott insulators'' in Fig. \ref{figure_triangle}. In the following, we will show that the first  BTI root state (the first row in Table \ref{table2})  can be obtained by pure $b\wedge \mathrm{d}a$ topological term where  U(1) charge symmetry is defined in an unusual way.

    \subsection{$\mathbb{Z}_2$ nature of   bulk U(1) symmetry definition}

   The unusual $U(1)$ symmetry transformation can be directly characterized by a   $\Theta$-term $F\wedge F$ in the response theory, where $F_{\mu\nu}$ is the field strength of external electromagnetic field $A_\mu$. Technically, one may start with a generic $b\wedge \text{d}a$ theory with $N$ components and then add an electromagnetic coupling terms like $\sum_I \frac{q_1^I}{4\pi}F_{\mu\nu}\partial_\lambda a^I_\rho\ep+\sum_I\frac{q_2^I}{4\pi} A_\mu \partial_\nu b^I_{\lambda\rho}\ep$ where $\{q_1^I\},\{q_2^I\}$ are two integral charge vectors. The first term is the coupling between face variable $F_{\mu\nu}$ and vortex-line current $\frac{1}{2\pi}\partial_\lambda a_\rho \ep$, where the additional $\frac{1}{2}$ is due to the double counting of the pair indices $\mu,\nu$. The second term is the coupling between link variable $A_\mu$ and boson particle current $\frac{1}{4\pi}\partial_\nu b_{\lambda\rho}\ep$.

 We may expect an electromagnetic response action with bulk $\Theta$-term in addition to the usual Maxwell terms:
\begin{align}
 \mathcal{S}_{\text{EM Response}}=\int d^4x \frac{\Theta}{32\pi^2 }F_{\mu\nu} F_{\lambda\rho}\ep+\cdots\,,\label{emthetaresponse}
\end{align}
where $\cdots$ denotes Maxwell terms. For  SPT states, there are only two choices: $
 \Theta=0\text{\,mod\,}4\pi$ or $2\pi \text{\,mod\,} 4\pi \,
$. This $\mathbb{Z}_2$ classification can be understood through alternative insights, such as the charge lattice of bulk quasiparticles\cite{3d5}, statistical Witten effect\cite{witten3}, both of which rely on the pioneering studies of dyon statistics in \cite{witten2}. But it is quite subtle if one proceeds the path-integral over $a^I, b^I$ to obtain the response action (\ref{emthetaresponse}) because $b^I$ and $a^I$ are constrained by several conditions. In the following,   We will  give a $N=2$ simple example in Sec. \ref{pbf2example}.

\subsection{An example with $N$=2}\label{pbf2example}


Let us start with a $b\wedge \mathrm{d}a$ theory with the following $K$ matrix:
 \begin{align}
K=\left(\begin{matrix}
      0 & 1\\
        1 & 2
            \end{matrix} \right)\,.\label{eqkmatrix}
\end{align}
More explicitly, the total Lagrangian is given by:
\begin{align}
\int d^4x\mathcal{L}_0=&\int i\frac{1}{2\pi} {b}^1\wedge \text{d} {a}^2 +i\int \frac{1}{2\pi} {b}^2\wedge \text{d}  {a}^1\nonumber\\
&+i\int \frac{2}{2\pi} {b}^2\wedge\text{d}{a}^2.\label{newlagr}
\end{align}
 Time-reversal symmetry is defined in the usual way shown in Eqs. (\ref{timerule0},\ref{timerule}). All gauge fields are constrained by Eqs. \ref{condition1_b} and \ref{condition1_a} (by adding indices $I,J,\cdots$).

 Let us move on to the surface theory. Starting with Eq. (\ref{newlagr}), one can proceed path-integral over  {compact} $a^1$ on an \textit{unoriented} manifold, leading to the  {quantization condition} on $b^2$:
 \begin{align}
 \varoiint_\mathcal{S} b^2=\pi\times \,\text{integer}\,\label{equation:higgsed_first_BTI}
 \end{align}
which indicates that $b^2$ is locally flat and   a new 1-form compact gauge field $\tilde{a}^2$ can be introduced in a way that is similar to Eq. (\ref{resolution_b}):
\begin{align}
  {b}^2_{\mu\nu}\xlongequal{\text{def.}}\partial_{[\mu} \tilde{a}^2_{\nu]}.\label{eq:b2a2}
 \end{align}
Thus, by means of Eqs. (\ref{timerule}), $\tilde{a}^2_\mu$ is transformed as a pseudo-vector under time-reversal symmetry:  $\tilde{a}^2_0\rightarrow -\tilde{a}^2_0, \tilde{a}^2_i\rightarrow  \tilde{a}^2_i$, $i=x,y,z$.

   The term $\sim  {b}^2\wedge \text{d}  {a}^2$ in Eq. (\ref{newlagr}) provides  a surface Chern-Simons term:
\begin{align}
\mathcal{L}_{\partial}=i\frac{1}{\pi}\epsilon^{\mu\nu\lambda} \tilde{a}^2_\mu\partial_\nu  {a}^2_\lambda\label{eq:toriccode}
\end{align}
which can also be reformulated by introducing a matrix $K_{\partial}\xlongequal{\text{def.}}\left(\begin{smallmatrix}
      0 & 2\\
        2 & 0
            \end{smallmatrix} \right)$ in the standard convention of $K$-matrix Chern-Simons theory.\cite{Wenbook} $ {a}^2_\mu$ and $\tilde{a}^2_\mu$ form a 2-dimensional vector $( {a}^2_\mu, \tilde{a}^2_\mu)^T$. The ground state of Eq. (\ref{eq:toriccode}) supports a Z$_2$ topological order\cite{kitaev1}
 associated with four gapped quasiparticle excitations ($1,e,m, \varepsilon$).  Here, ``$1$'' denotes identical particles. Both $e$ and $m$ are bosonic, while $\varepsilon$ is a fermion. The particle $e$   carries $+1$ gauge charge of $ {a}^2_\mu$, while, the particle $m$ carries $+1$ gauge charge of $ \tilde{a}^2_\mu$. $\varepsilon$ carries $+1$ gauge charges of both gauge fields. These three nontrivial quasiparticles all have mutual semionic statistics,  \textit{i.e.}   full braiding one particle around another distinct particle leads to a $\pi$ Aharonov-Bohm phase.

To obtain BTI states, we expect the U(1) symmetry transformation may be performed in an unusual way. For the purpose, we may add  the  following  electromagnetic coupling term to $\mathcal{L}_0$ in Eq. (\ref{newlagr}):
  \begin{align}
   \frac{q_1}{4\pi}F_{\mu\nu}\partial_\lambda  {a}^2_\rho \ep +\frac{q_2}{4\pi}A_{\mu}  \partial_\nu   {b}^2_{\lambda\rho}\ep\,.\label{eq:3dresponse}
\end{align}
%
Let us consider that a surface  is located on $z=0$ plane. The surface theory is described by Eq. (\ref{eq:toriccode}). The coupling term contributes the following surface electromagnetic coupling terms:
  \begin{align}
  \frac{q_1}{2\pi}\epsilon^{\mu\nu\lambda}A_{\mu}\partial_\nu  {a}^2_\lambda  +\frac{q_2}{2\pi}\epsilon^{\mu\nu\lambda}A_\mu \partial_\nu \tilde{a}^2_{\lambda}\,,\label{eq:3dresponse2d}
 \end{align}
 where Eq. (\ref{eq:b2a2}) is applied.
Based on the Chern-Simons term in Eq. (\ref{eq:toriccode}),  one may calculate the electric charge  carried by each quasiparticle:
\begin{align}
&Q_e=(q_2,q_1)(K_\partial)^{-1}(1,0)^T=\frac{q_1}{2}\,,\nonumber\\
&Q_m=(q_2,q_1)(K_\partial)^{-1}(0,1)^T=\frac{q_2}{2}\,.\nonumber
\end{align}
 Physically, both $e$ and $m$ quasiparticles can always attach trivial identity particles to change their charges by arbitrary integer so that $q_1$ and $q_2$ are integers mod $2$, namely,
\begin{align}
q_1\sim q_1+2\,,q_2\sim q_2+2\,.\label{qq}
\end{align}
  For this reason, it is sufficient to merely consider  the following four choices: $(q_2,q_1)=(0,0),(0,1),(1,0),(1,1)$. The first three choices can be realized on a 2d plane without breaking time-reversal symmetry since   the Hall conductance $\sigma_{xy}=\frac{1}{2\pi}q^T(K_\partial)^{-1}q=\frac{q_1q_2}{2\pi}=0$ on a 2d plane.   However $(1,1)$ necessarily breaks time-reversal symmetry on a 2d plane since its Hall conductance $\sigma_{xy}=\frac{1}{2\pi}q^T(K_\partial)^{-1}q=\frac{1}{2\pi}q_1q_2=\frac{1}{2\pi}$ is nonzero. Thus, there is a nonvanishing chiral charge flow on the 1d edge of the 2d plane. However, it doesn't break time-reversal symmetry on the surface since chiral charge flow is not a   physical observable on the surface.  Therefore, the charge assignment $(1,1)$ faces obstruction in being realized on a 2d plane with U(1)$\rtimes$Z$^T_2$ symmetry and this obstruction leads to a BTI in which both $e$ and $m$ carry half-charge on the surface.   Note that, the $q_1$-coupling terms in  Eq. (\ref{eq:3dresponse}) and Eq. (\ref{eq:3dresponse2d}) change sign under time-reversal transformation. However, the sign can be removed through shifting $q_1$ to $-q_1$  following the identification in Eq. (\ref{qq}).

In a real transport experiment,  one may explicitly break time-reversal symmetry along the normal direction on the surface with charge assignment $(1,1)$ and put the 3d system in a slab geometry.  
The surface quantum Hall conductance is quantized at $\frac{1}{2\pi}$, which corresponds to a surface response action in the form of Chern-Simons term: $\frac{1}{4\pi}A_\mu\partial_\nu A_\lambda\epsilon^{\mu\nu\lambda}$. It can be formally extended to a bulk $\Theta$-term, \textit{i.e.} Eq. (\ref{emthetaresponse}) with $\Theta$ angle given by $
 \Theta=2\pi \text{\,mod\,} 4\pi \,
$ and the following generic relation:
\begin{align}
\Theta=4\pi^2\sigma_{xy}.\label{thetasigma}
\end{align}
The $4\pi$ periodicity corresponds to even-quantized Hall conductance $\sigma_{xy}=\frac{1}{2\pi}\times2k$, which can be realized in purely 2D bosonic systems.
A projective construction on such a BTI with $\Theta$ response has been made in details in Ref. \onlinecite{3d5}.

Physically, $e$ and $m$ particles can be regarded as ends of condensed vortex-lines here, and we may think these invisible
vortex-lines carry integer charge and form a nontrivial 1D BTI phase. Thus, it is also not a surprise that the end of these vortex-lines will carry half-charge.

\subsection{Many-to-one correspondence between surface and bulk}\label{many}
The above discussion illustrates how the surface Z$_2$ topological order arise with unusual U(1) symmetry transformations.  As a matter of fact, the surface may be of different kinds while all share the same bulk that  only supports trivial boson excitations.  In other words, the surface topological order may be much richer.

One may replace the $K$-matrix in Eq. (\ref{eqkmatrix}) by
\begin{align}
K=\left(\begin{matrix}
      0 & 1\\
        1 & p
            \end{matrix} \right)\,,\label{eqkmatrixgeneral}
\end{align}
where $p$ is a nonzero positive integer. The surface term Eq. (\ref{eq:toriccode}) now becomes:
\begin{align}
\mathcal{L}_\partial=i\frac{p}{2\pi}\epsilon^{\mu\nu\lambda} \tilde{a}^2_\mu\partial_\nu  {a}^2_\lambda\label{newt}
\end{align}
 which corresponds to a Z$_p$ topological order on the surface labeled by $K_\partial=\left(\begin{smallmatrix}
      0 & p\\
        p & 0
            \end{smallmatrix} \right)$.

  There are $p^2$ types of quasiparticles (including the trivial particle), which are labeled by quasiparticle vector $l=(l_1,l_2)^T$ with $l_1,l_2=0,1,\cdots,p-1$. The $l^{\rm th}$-quasiparticle carries the electric charge \begin{align}
  Q_l=q^TK^{-1}_\partial l=\frac{1}{p}(q_1l_2+q_2l_1).\label{equationc}
  \end{align}
   Physically, all quasiparticles can attach trivial identity particles such that their charges can be changed by any integer. Therefore, the following identification conditions exist:
\begin{align}
q_1\sim q_1+p\,,\,q_2\sim q_2+p\,.\label{qq1}
\end{align}

 We note that $q_1$-coupling terms in  Eq. (\ref{eq:3dresponse}) and Eq. (\ref{eq:3dresponse2d}) change sign under time-reversal transformation. However, the sign can be removed through shifting $q_1$ to $-q_1$  following the identification in Eq. (\ref{qq1}).  Therefore, bulk time-reversal symmetry requires that $p=\text{even}$ and only two choices for the integer $q_1$ are legitimate: $q_1=0\text{mod}p,\, \frac{p}{2}\text{mod}p$.

To determine the bulk is trivial or not, one may again examine the surface Hall conductance which  is quantized at odd for nontrivial bulk. All even-quantized parts can be removed by stacking several U(1) SPT states on the surface.   At present, the Hall conductance is given by $\sigma_{xy}=\frac{1}{2\pi}q^TK^{-1}_\partial q=\frac{1}{2\pi}\frac{2q_1q_2}{p}$.  Combined with the relation in Eq. (\ref{thetasigma}), we end up with Table \ref{table3}.
  \begin{table}
\caption{The surface of the first BTI root states (shown in the first row) labeled by three integers ($p,q_1,q_2$).}\label{table3}
 \begin{tabular}[t]{cccc}
\hline\hline
 \begin{minipage}[t]{0.5in}$p$\end{minipage} &\begin{minipage}[t]{1.5in}  $q_1$, $q_2$  \end{minipage}& \begin{minipage}[t]{1in}    Bulk \end{minipage}  \\\hline \\
 \begin{minipage}[t]{0.5in}$p=$even\end{minipage} &\begin{minipage}[t]{1.5in}  $q_1=\frac{p}{2}\text{mod}p$, $q_2=\text{odd}$ \end{minipage}& \begin{minipage}[t]{1in}The first BTI root states  \end{minipage}
 ~\\   \\
 \begin{minipage}[t]{0.5in}$p=$even\end{minipage} &\begin{minipage}[t]{1.5in}  $q_1=\frac{p}{2}\text{mod}p$, $q_2=\text{even}$ \end{minipage}& \begin{minipage}[t]{1in}trivial states  \end{minipage}
 ~\\  \\
 \begin{minipage}[t]{0.5in}$p=$even\end{minipage} &\begin{minipage}[t]{1.5in}   $q_1=0\text{mod}p$, $q_2=$any integers\end{minipage}& \begin{minipage}[t]{1in}trivial states  \end{minipage}
 ~\\  \\
 \begin{minipage}[t]{0.5in}$p=$even\end{minipage} &\begin{minipage}[t]{1.5in}  $q_1\neq0\text{mod}p$, $q_1\neq \frac{p}{2}\text{mod}p$, $q_2=$any integers \end{minipage}& \begin{minipage}[t]{1in}  Z$^T_2$ is   broken.  \end{minipage}
 ~\\  \\
 \begin{minipage}[t]{0.5in}$p=$odd\end{minipage} &\begin{minipage}[t]{1.5in}  $q_1, q_2$=any integers  \end{minipage}& \begin{minipage}[t]{1in}  Z$^T_2$ is   broken.\end{minipage}
 ~\\
      \hline\hline
  \end{tabular}
  \end{table}
%
As a side note, in the first two cases,  by means of the identification in Eq. (\ref{qq1}), $q_2$ may be shifted by $p$. Since $p$ is even in these two cases, the even / odd property of $q_2$ is unchanged. Thus, these two cases are consistent with the identification conditions.

When $p=2,q_1=q_2=1$, the theory goes back to Sec. \ref{pbf2example}. It is clear that  the surface Z$_2$ topological order obtained in Sec. \ref{pbf2example} is just one possible surface of BTI states, which manifests the physics of many-to-one correspondence between surface and bulk.
For example, we may choose $p=4,q_1=2,q_2=1$ such that the bulk is a BTI state with nontrivial Witten effect.
The electric charge carried by totally $4^2-1=15$ nontrivial quasiparticles can be calculated by Eq. (\ref{equationc}): $Q_l=\frac{l_2}{2}+\frac{l_1}{4}$. Such an assignment of fractional charge on quasiparticles of Z$_4$ topological order cannot be realized on a 2d plane unless breaking time-reversal symmetry.

    \section{Bosonic topological insulators from pure $b\wedge \mathrm{d}a$ term$-$(II)}\label{sec:pbf1}
   In Sec. \ref{sec:pbf2}, we have shown that the first  BTI root state  can be obtained by a pure $b\wedge \mathrm{d}a$ topological term where  U(1) charge symmetry is defined in an unusual way.
 In the following, we will continue to show that the second  BTI root state (the second row in Table \ref{table2})  can be obtained by a pure $b\wedge \mathrm{d}a$ topological term where time-reversal symmetry is defined in an unusual way.

   \subsection{$\mathbb{Z}_2$ nature of bulk time-reversal symmetry definition}

Let us   consider $K$ matrix in the form of Eq. (\ref{eq:K}). A time reversal transformation acting on   gauge fields $a^I_\mu,b^{I}_{\mu\nu}$ can be formally expressed as:
\begin{align}
\mathcal{T}{a}^I_0\mathcal{T}^{-1}=T^a_{IJ}a^J_0\,\,,\,\,\mathcal{T}{b}^I_{0i}\mathcal{T}^{-1}=T^b_{IJ}b^J_{0i}\,\,,\,\label{eq:etplus}\\
\mathcal{T}{a}^I_{i}\mathcal{T}^{-1}=-T^a_{IJ}a^J_{i}\,\,,\,\,\mathcal{T}{b}^I_{ij}\mathcal{T}^{-1}=-T^b_{IJ}b^J_{ij}\,,\label{eq:et}
\end{align}
where $T^a$ and $T^b$ are two integer-valued matrices. In the following, we will simply call $T^a$ and $T^b$ ``$T$-matrices''.  After transforming twice, all gauge variables are unchanged. So we have the constraint $(T^a)^2=(T^b)^2=1$. It indicates that $|\text{det}T^a|=|\text{det}T^b|=1$ and both matrices belong to a subset of $\mathbb{GL}(N,\mathbb{Z}$) group.

After $\mathbb{GL}$ transformations and time-reversal transformation, $K$ matrix is transformed to a new one but $|\text{det}K|=1$ is still valid such that the bulk still merely supports trivial gapped boson excitations as before. From this perspective, the bulk always keeps time-reversal symmetry although the formal expression of Lagrangian is given by a new $K$ matrix.

On the other hand, one may apply arbitrary $\mathbb{GL}$($N$,$\mathbb{Z}$) transformations on both sides of all equations in Eqs. (\ref{eq:etplus},\ref{eq:et}). Using the notation in Eq. (\ref{eq:gltrans}), we obtain the following equations:
\begin{align}
\mathcal{T}\underline{{a}}^I_0\mathcal{T}^{-1}= \underline{T}^a_{IJ} \underline{a}^J_0\,\,,\,\,\mathcal{T}\underline{{b}}^I_{0i}\mathcal{T}^{-1}= \underline{T}^b_{IJ}  \underline{b}^J_{0i}\,\,,\, \\
\mathcal{T}\underline{{a}}^I_{i}\mathcal{T}^{-1}=-\underline{T}^a_{IJ} \underline{a}^J_{i}\,\,,\,\,\mathcal{T}\underline{{b}}^I_{ij}\mathcal{T}^{-1}=-\underline{T}^b_{IJ}\underline{b}^J_{ij}\,,\label{eq:et1}
\end{align}
where two $T$ matrices are transformed to two new ones:
\begin{align}
\underline{T}^a\xlongequal{\text{def.}}M^{-1}T^a M\,\,,\,\underline{T}^b\xlongequal{\text{def.}}W^{-1}T^b W\,\,.\label{equation:operatortr}
\end{align}

By keeping  $|\text{det}W|=|\text{det}M|=1$ in mind, it is clear that the $\pm$ sign of determinant of $T$ matrices is manifestly invariant under arbitrary sequence of formal $\mathbb{GL}$ transformations. For the sake of convenience, let us introduce a notation $(a,b)$ that denotes such $\pm$ signs of determinants:
\begin{align}
(a,b)\xlongequal{\text{def.}} (\text{sign of det$T^a$}\,,\text{ sign of det$T^b$})\,.\nonumber
\end{align}
 Due to the presence of this invariant, we are able to understand usual time-reversal transformation defined in Eqs. (\ref{timerule0}) and (\ref{timerule}) in a much more general background. In terms of $T$ matrices, the usual time-reversal transformation defined in Eqs. (\ref{timerule0}) and (\ref{timerule}) is denoted by:
\begin{align}
T^a=-T^b=\text{diag}(1,1,\cdots,1)_{N\times N}\,,\label{equation:trsymmetrytrivial}
\end{align}
where $K$ matrix is fixed as a unit matrix shown in Eq. (\ref{eq:K}). This specific form of usual time-reversal transformation in a given basis (i.e. $K=\mathbb{I}$) can be generalized and replaced by the following invariant:
 \begin{align}
(a,b)=(1,(-1)^N)\label{eq:abinvariant}
\end{align}
which is a universal property of all specific forms of usual time-reversal transformations.

To sum up, let us consider a 3d bulk state described by a $N$-component $b\wedge \mathrm{d}a$ term labeled by $K$ with $|\text{det}K|=1$. There are $T^a$ and $T^b$ two matrices that define time-reversal transformations.  As discussed before, for the purpose of exploring nontrivial SPT states, we consider the cases that have symmetry-preserving surface. Then, if $(a,b)=(1,(-1)^N)$, the state admits a usual time-reversal transformation and thereby a trivial SPT state.
 If $(a,b)\neq(1,(-1)^{N})$, the state is a nontrivial SPT state.  From this point of view,  a $\mathbb{Z}_2$ classification is obtained by attempting to change the definition of time-reversal transformations.  Along this line of thinking, in Sec. \ref{sec:purebftime}, we will study $N=2$ as a simple example which reproduces the BTI state labeled by the first $\mathbb{Z}_2$ index introduced in Sec. \ref{sec:introduction}.

 \subsection{An example with $N=2$}\label{sec:purebftime}

Let us still start with a $b\wedge \mathrm{d}a$ theory with     $K$ matrix given by (\ref{eqkmatrix}). The total Lagrangian is given by (\ref{newlagr}).  However, at   present, the time-reversal transformation  is defined as
\begin{align}
&\mathcal{T}{ {a}}^1_0\mathcal{T}^{-1}=  {a}^1_0\,\,,\mathcal{T}{ {a}}^1_i\mathcal{T}^{-1} = -{a}^1_i \,,\label{tr1} \\
&\mathcal{T}{ {b}}^{1}_{0,i}\mathcal{T}^{-1}={b}^{1}_{0,i}\,\,,\mathcal{T}{{b}}^{1}_{i,j}\mathcal{T}^{-1}= -\underline{b}^{1}_{i,j}\,\, ,\,\label{timerule1} \\
&\mathcal{T}{ {a}}^2_0\mathcal{T}^{-1}= - {a}^2_0\,\,,\mathcal{T}{ {a}}^2_i\mathcal{T}^{-1}=   {a}^2_i  \,,\label{a2time}\\
&\mathcal{T}{ {b}}^{2}_{0,i} \mathcal{T}^{-1}= - {b}^{2}_{0,i}\,\,,\mathcal{T}{ {b}}^{2}_{i,j}\mathcal{T}^{-1}=  {b}^{2}_{i,j}\,\, .\,\label{timerule2}
\end{align}
which is different from the usual definition Eqs. (\ref{timerule0}) and (\ref{timerule}).
The associated $ {T}$ matrices of the time-reversal operator $\mathcal{T}$ are given by:
 \begin{align}
 {T}^a=\left(\begin{matrix}
      1 & 0\\
        0 & -1
            \end{matrix} \right)\,\,,\, {T}^b=\left(\begin{matrix}
      1 & 0\\
        0 & -1
            \end{matrix} \right)\,.\label{newTmatrix}
\end{align}
This definition of time-reversal symmetry transformation is labeled by $(a,b)=(-1,-1)$ which is nontrivial according to Eq. (\ref{eq:abinvariant}).
Under the time-reversal transformation, the $b\wedge \mathrm{d}a$ term labeled by $K$ is transformed to the term labeled by
  $K'=\left(\begin{smallmatrix}
      0 & 1\\
        1 & -2
            \end{smallmatrix} \right)
$. At first glance, time-reversal symmetry is broken in the bulk. However, two $b\wedge\mathrm{d}a$ theories labeled by $K'$ and $K$ are $\mathbb{GL}$-equivalent by using
\begin{align}
W=\left(\begin{matrix}
      1 & 0\\
        0 & -1
            \end{matrix} \right)\,,\,M=\left(\begin{matrix}
     - 1 & 0\\
        0 & 1
            \end{matrix} \right)\nonumber
\end{align}
defined in Eq. (\ref{eq:wm}). More explicitly, these $\mathbb{GL}$  transformations lead to a sign change in both $b^2_{\mu\nu}$ and $a^1_\mu$ while $b^1_{\mu\nu}$ and $a^2_\mu$ are invariant:
\begin{align}
&b^2_{\mu\nu}\rightarrow -b^2_{\mu\nu}\,,\,a^1_{\mu}\rightarrow -a^1_{\mu}\,,\label{gltransfbb1}\\
&b^1_{\mu\nu}\rightarrow b^1_{\mu\nu}\,,\,a^2_{\mu}\rightarrow a^2_{\mu}\,\,.\label{gltransfbb2}
\end{align}

 Let us move on to the surface theory (\ref{eq:toriccode}).   According to the transformation of $ {b}^2_{\mu\nu}$ in Eq. (\ref{timerule2}), the time-reversal transformation rule of $\tilde{a}^2_\mu$ defined in Eq. (\ref{eq:b2a2}) is automatically fixed:   \begin{align}
\mathcal{T}{\tilde{a}}^2_0\mathcal{T}^{-1}=-\tilde{a}^2_0\,\,,\,\mathcal{T}{\tilde{a}}^2_i\mathcal{T}^{-1}= \tilde{a}^2_i  \,.\label{eq:newtransform}
\end{align}

  Due to the time-reversal transformations in Eqs. (\ref{eq:newtransform},\ref{a2time}),  both $m$ and $e$, which carry unit gauge charges of $\tilde{a}^2_\mu$ and $a^2_\mu$ respectively, are pseudo-like particles on the surface. Under these time-reversal transformations,  a minus sign appears in the Chern-Simons term in Eq. (\ref{eq:toriccode}). Despite that, the surface state doesn't break time-reversal.  More precisely, the appearance of this minus sign  leaves the set of surface physical observables  in Definition \ref{remarkobs} unaffected by noting that there is a $\mathbb{GL}$ equivalence relation: $P^T(-K_{\partial})  P=K_{\partial}$, where $P=\left(\begin{smallmatrix}
      1 & 0\\
        0 & -1
            \end{smallmatrix} \right)$.  Therefore, the surface is symmetric under time-reversal transformation.

             It is also beneficial to investigate the equations of motion (EOM) of gauge fields under time-reversal symmetry. By adding two quasiparticle currents $j^e_\mu a^2_\mu+j^m_\mu \tilde{a}^2_\mu$ where $j^e_\mu$ and $j^m_\mu$ are $e$-particle and $m$-particle currents respectively, the EOMs of gauge fields are given by (only zero-component is shown here without loss of generality):
$           \pi \rho^e= \nabla \times \tilde{\mathbf{a}}^2\xlongequal{\text{def.}}\tilde{\mathcal{B}} \,,\,           \pi \rho^m=\nabla \times  {\mathbf{a}}^2\xlongequal{\text{def.}}{\mathcal{B}}
  $, where $\rho^e=j^e_0$ and $\rho^m=j^m_0$ are density variables of $e$ and $m$ respectively. $\tilde{\mathcal{B}}$ and $\mathcal{B}$ are magnetic flux strength of $\tilde{\mathbf{a}}^2$ and $\mathbf{a}^2$, respectively. Under time-reversal, $\rho_e$ and $\rho_m$ change sign since they are pseudo-like: $\rho_e\rightarrow -\rho_e\,,\,\rho_m\rightarrow -\rho_m$. On the other hand, both  $\tilde{\mathcal{B}}$ and $\mathcal{B}$  are unchanged under time-reversal transformations Eqs. (\ref{eq:newtransform},\ref{a2time}). It seems that EOMs break time-reversal. However time-reversal symmetry is still unbroken due to the very existence of the 3d bulk.  More concretely, the 3d bulk can source a trivial particle that carries two units of $a^2_\mu$ gauge charge. One may attach this trivial particle to a time-reversal partner of $e$ particles on the surface, rendering $-\rho_e+2\rho_e=\rho_e$. Likewise, the 3d bulk can source  a trivial particle that carries two units of $\tilde{a}^2_\mu$ gauge charge. One may attach this trivial particle to a time-reversal partner of $m$ particles,  rendering $-\rho_m+2\rho_m=\rho_m$. As a result, both EOMs respect time-reversal symmetry.

 However, this Z$_2$ topological order state on a 2d plane (i.e. no 3d bulk) necessarily breaks time-reversal symmetry. In the absence of 3d bulk, all trivial particles that are used to change sign of $\rho_e$ and $\rho_m$ can only come from 2d state itself. Consequently the magnetic fluxes generated by these trivial particles will change $\int dxdy\tilde{\mathcal{B}}$ and $\int dxdy{\mathcal{B}}$  to $\int dxdy\tilde{\mathcal{B}}+2\pi$ and $\int dxdy\mathcal{B}+2\pi$ respectively.  Thus, EOMs always break time-reversal symmetry.

  In summary,   following the definition of obstruction in Definition \ref{definition2dplane}, the 3d bulk is a nontrivial SPT state, i.e. a BTI state.
Physically, the unique way to realize such a time-reversal symmetry on gauge fields, i.e. Eqs. (\ref{eq:newtransform},\ref{a2time}), is to consider that both $e$ and $m$ are Kramers' doublets,  {and $\rho_e,\rho_m$ should be regarded as spin density $S_z$ of a spin-$1/2$ particle. We note that by "spin-$1/2$" here, we really mean a projective representation of time reversal symmetry and it has nothing to do with spin rotational symmetry. In the bulk, all particles must carry linear representation of time reversal symmetry, therefore, the spin-$1/2$ particle on the surface can not be screened.} Other possibilities of Kramers' degeneracy assignment (e.g. $e$ is Kramers' doublet while $m$ is Kramers' singlet) can be realized on a 2d plane without breaking time-reversal symmetry.\cite{levinstern,3d4}
   This obstruction provides us a physical way to understand the nontrivial BTI root phase generated by exotic time reversal symmetry.   Indeed, both $e$ and $m$ particles can be regarded as ends of vortex-lines that are condensed and invisible in the bulk. In terms of simple physical picture, we may think these invisible
vortex-lines carry integer spin and form a nontrivial 1D SPT phase, e.g., the Haldane phase. Therefore, it is not a surprise that the end of these vortex-lines carry half-integer spins which form Kramers' doublets under time reversal symmetry.

\section{Z$_N$ SPT in three dimensions: beyond group cohomology theory}\label{sec:znspt}
In this section, we use the one component action Eq.(\ref{eq:LBH}) to discuss possible Z$_N$ symmetry protected phases beyond group cohomology class.
Let us assume a generic Z$_N$ SPT in 3d can be described by:
\begin{align}
\mathcal{L}=i \frac{1}{4\pi}a_{\mu}\partial_{\nu} b_{\lambda\rho}\ep +i\frac{\Lambda}{16\pi}b_{\mu\nu}b_{\lambda\rho}\ep\,,\nonumber
\end{align}
where only one-component is taken into account for convenience. Here, $a_{\mu}$ and $b_{\mu\nu}$ are still non-compact and compact respectively. In contrast to the previous discussion of U(1)$\rtimes$Z$^T_2$ where non-vanishing quantized $\Lambda$ needs the help of time-reversal symmetry, $\Lambda$ in Z$_N$ SPT is supposed to be quantized even without the help of time-reversal symmetry. Following \cite{2d10}, let us consider the following gauge coupling to  ``probe the Z$_N$ SPT order'':
\begin{align}
\mathcal{L}_{coupling}=i\frac{1}{4\pi}B_{\mu\nu}\partial_{\lambda}a_{\rho}\ep+i\frac{N}{4\pi}B_{\mu\nu}\partial_\lambda A_{\rho}\ep\,.\nonumber
\end{align}
Several explanations are in order. First, the $B_{\mu\nu}$ gauge field in $B\wedge \text{d}a$ term is a ``2-form compact probe field'' that minimally couples to strings ($2\pi$-vortex-lines in the ground state). It is this type of coupling that is missed in Dijkgraaf-Witten gauge theory\cite{DWmodel} since $H^4[\text{Z}_N,U(1)]=\mathbb{Z}_1$.  The term $B\wedge \text{d}A$ can be viewed as a Higgs condensate term\cite{2d10} where $A_{\mu}$ is a 1-form \textit{compact} gauge field introduced by Hubbard-Stratonovich transformation. By means of this term, the probe field $B_{\mu\nu}$ is naturally higgssed to Z$_N$ discrete gauge field.

Now, we are in a position to integrate out all SPT degrees of freedom. Integrating the non-compact field $a_{\mu}$ renders $b_{\mu\nu}=B_{\mu\nu}$.  Consequently, we end up with an action in the background fields $B$ and $A$:\cite{wedge_convention}
\begin{align}
 S=&i\int d^4x\frac{N}{4\pi}B_{\mu\nu}\partial_\lambda A_{\rho}\ep+i\int d^4x\frac{\Lambda }{16\pi}B_{\mu\nu} B_{\lambda\rho}\ep\nonumber\\
=&i\frac{N}{2\pi}\int B\wedge \mathrm{d} A+i\frac{\Lambda }{4\pi} \int B\wedge B\,,\label{equation:zntheory}
\end{align}
All possible values of $\Lambda$ can be found by using the procedures in Sec. \ref{sec:periodicity}. Finally, we end up with the following quantization condition (more details are present in Appendix \ref{derivingequation:quantizationzn}):
 \begin{align}
 \Lambda/N\in \mathbb{Z}\,.\label{equation:quantizationzn}
 \end{align}
 The   constraints (\ref{equation:periodicityzn}) and (\ref{equation:quantizationzn}) suggest Z$_N$ different SPT states protected by Z$_N$ symmetries in three dimensions.

 However, since the probe field here is a two-form gauge field with a flat connection, it cannot be regarded as the symmetry twist of a usual condensed matter system where symmetry charges are carried by point-like particles. Instead, we need to consider a system consisting of string-like objects carrying global symmetry quantum numbers, for example, in the context of "Generalized Global Symmetries" in string theory by Davide \emph{etal}\cite{Davide} recently. Although for solid state systems, we are not aware how to prepare string-like objets carrying global quantum numbers, we hope that certain artificial quantum systems, e.g., cold atom systems, might be able to realize these exotic quantum phases.

 \section{Conclusions}\label{sec:con}
		

  In this paper, based on a physical process called ``vortex-line condensation'' in three-dimensional superfluids, we have  constructed a bulk dynamical TQFT description Eq. (\ref{action}) of all three bosonic topological insulator states (BTI).  The schematic phase diagram is shown in Fig. \ref{figure_triangle}.  Such a physical way of thinking allows us to understand the physical meaning of each gauge field variable, and, most importantly,   symmetry definitions in the bulk.  Our method  will further shed light on a more challenging question: how to design microscopic interactions to realize BTI states in solid state materials or ultra-cold-atom experiments? Especially, it is quite interesting to explore the possible interaction terms to realize the linking Berry phase contributed by $b\wedge b$ as shown in Fig. \ref{figure_sheet}. For those two BTI states (Sec. \ref{sec:pbf2}, \ref{sec:pbf1}) that do not need $b\wedge b$ term, one may attach either charge or spin degrees of freedom to vortex-lines, which may result in BTI states after vortex-lines condense.

   We have shown that one of the three BTI states requires a nontrivial existence of $b\wedge b$ term, which is beyond group cohomology theory.  The remaining two states are within group cohomology theory and can be constructed from pure $b\wedge \mathrm{d}a$ term. In contrast to previous works of BTI where the surface topological order is always Z$_2$ topological order, now, the surface topological order of  the BTI state within group cohomology classification can be a generic Z$_p$ topological order with even $p$ as shown in Sec. \ref{sec:pbf2}. This many-to-one correspondence between surface and bulk explicitly indicates the knowledge about bulk field theory is highly desirable, which is also explained in details in Sec. \ref{sec:introduction}. In addition to BTI, applying our construction of BTI to 3d SPT with Z$_N$ unitary symmetry suggests some non-trivial Z$_N$
    SPT state in three dimensions beyond the group cohomology classification.   However, in contrast to the usual quantum systems where global symmetry quantum number is carried by point-like particles, the new class of Z$_N$ SPT phases proposed here requires string-like object to carry the global symmetry quantum number. Finally, based on the results we obtained, we conjecture that all SPT phases described by  Eq. (\ref{action}) with a nontrivial $b\wedge b$ term are generally beyond the group cohomology classification.

    In the future, it will be interesting to apply such a physical derivation of bulk dynamical TQFT to other SPT states, even including fermionic SPT states where a spin manifold is required. A challenging  problem is the bulk dynamical TQFT(not response theory) description of FTI both in free-fermionic\cite{TI1,TI2,TI3,TI4,TI5,TI6,TIexp} and interacting cases\cite{interacting1,interacting2,interacting3,interacting4}. There are many previous important efforts, such as \cite{Cho,atma}. In \cite{atma1},  functional bosonization techniques are applied and  $b\wedge b$ term appears.  We believe that the basic methodology presented in our work combined with the previous efforts will shed light on this hard problem.

    \section*{Acknowledgements}
We would like to thank Meng  Cheng,   Chao-Ming Jian,   Joseph Maciejko,     Xiao-Gang Wen, and Shing-Tung Yau for enlightening discussions and/or comments. We thank Ching Hua Lee and Gang Chen for    many valuable suggestions on the manuscript. We also thank Princeton Center for Theoretical Science at Princeton University for hospitality during the ``\textit{Symmetry in Topological Phases}'' Workshop (Mar. 2014) where this work was initiated. P.Y. especially acknowledges  Professor Zheng-Yu Weng's host at Institute for Advanced Study in Tsinghua University  in Beijing (Jul. 2014) where the work was done in part. Research at Perimeter Institute is supported by the Government of Canada through Industry Canada and by the Province of Ontario through the Ministry of Economic Development \& Innovation.

   \appendix

 \section{Quantization conditions and evaluations of partition function on closed manifolds}\label{appendix_gsd}
 In Sec. \ref{sec:periodicity}, we obtained the  {quantization condition} Eq. (\ref{equation:higgsed}) based on the physical approach, namely, the microscopic origin `vortex-line condensation'.  Although in the whole derivation of the third BTI root state (see Table \ref{table2}) we will consider this physical approach and $a$ is noncompact,  in this Appendix, we shall derive the same quantization condition in a mathematical setting that is  independent of the microscopic origin. First of all, we may assume that both $b$ and $a$ satisfy the following compactification conditions:
\begin{align}
&\int\!\!\!\!\!\int \!\!\!\!\! \int_{\mathcal{V}'} \text{d}b=2\pi\times \text{integer}\,,\label{condition1_b}\\
&\varoiint_\mathcal{S} \text{d}a=2\pi\times \text{integer}\,,\label{condition1_a}
 \end{align}
 where, both 3d manifold $\mathcal{V}'$ and 2d manifold $\mathcal{S}$ are closed.  Let us consider the partition function:
 \begin{align}
 \mathsf{Z}=\int \mathscr{D}[a]\mathscr{D}[b]e^{- S}\,,
 \end{align}
 where the classical action $S$ is given by Eq. (\ref{equation:onecomponenttheory}). Integration over $\text{d}a$ is a Poisson summation which directly  leads to the  {quantization condition} on $b$ given by Eq. (\ref{equation:higgsed}). Therefore, both ways give the same answer. The treatment in the main text can be viewed as a semiclassical way, while, the new way is more rigorous since the nontrivial shift in Eq. (\ref{gaugetransform0}) generally renders compactification of $a$ once $b$ is assumed to be compact. A similar line of thinking is recently given in Ref. \cite{gupaper}.

 We can also check whether the addition of $b\wedge b$ term will induce fractionalization, namely topological order in the bulk. Following the spirit of Laughlin thought experiment for probing electron fractionalization in  fractional quantum Hall effects,   we may apply an external gauge field that minimally couples to the conserved currents of matter fields, i.e. bosonic point-particles and strings (vortex-lines). In the present 3d state, we add a 2-form external gauge field $B$ through the minimal coupling term ``$\frac{1}{2\pi}B\wedge \text{d}a$'' where ``$\frac{1}{2\pi} ^*\text{d}a$'' denotes the 2-form string (vortex-line) current. ``$^*$'' denotes Hodge dual. We choose $B$ instead of the usual 1-form external electromagnetic field $A$ to probe fractionalization for the reason that in 3d state, anyons do not exist while possible fractionalization only comes from fractionalization of flux strength of vortex-lines, namely, $2\pi$ becomes $2\pi/k$ with $k>1,k\in\mathbb{Z}$.

  Then the GSD (ground state degeneracy) ratio in a given spacetime manifold between the theories with and without $b\wedge b$ is given by the ratio of partition functions in the presence of external gauge field $B$:
 \begin{align}
\frac{\mathsf{Z}}{\mathsf{Z}_0}=\frac{1}{\mathsf{Z}_0} \int \mathscr{D}[a]\mathscr{D}[b]e^{-i\int( \frac{1}{2\pi}b\wedge \text{d}a+\frac{\Lambda}{4\pi} b\wedge b-\frac{1}{2\pi}a\wedge \text{d}B)}
 \end{align}
 where, $\mathsf{Z}_0$ is given by: $\mathsf{Z}_0=\mathsf{Z}[\Lambda=0]$ by definition.
 Integration over $a$ is a Poisson summation which leads to
  \begin{align}
 \varoiint_\mathcal{S} (b+B)=2\pi\times \,\text{integer}\,,\label{equation:higgsed1234}
 \end{align}
 where, $\mathcal{S}$  is a closed 2d manifold forming a surface of a 3d space $\mathcal{V}$, i.e. $\mathcal{S}=\partial\mathcal{V}$.
 
 We note that the probe field $B$ satisfies the  {quantization condition}:
  \begin{align}
 \varoiint_\mathcal{S} B=2\pi\times \,\text{integer}\,.\label{equation:higgsed12345}
 \end{align}
Thus we may set $B=b$ up to gauge transformation (including large one) in $b\wedge b$ term, which leads to the following result:
  \begin{align}
\frac{\mathsf{Z}}{\mathsf{Z}_0}=e^{i\int\frac{\Lambda}{4\pi} B\wedge B}
 \end{align}
 In spacetime with different topology, we may calculate the ratio that is always a phase factor with unit length, i.e. $\left|\frac{\mathsf{Z}}{\mathsf{Z}_0}\right|=1$. For example, one may choose $\mathbb{T}_{0x}\times \mathbb{T}_{yz}$ topology and calculate the ratio. The calculation is completely same as Appendix \ref{sub1} by just replacing $b$ by $B$.
  In summary,  the GSD ratio in the presence of $b\wedge b$ is still one in all kinds of spacetime topology. Since the pure $b\wedge \text{d}a$ theory is at level-$1$ with GSD=1. Thus, the addition of $b\wedge b$ doesn't induce new GSD and thus fractionalization in our theory.

 \section{Some technical details}\label{derivingequation:quantizationzn}

\subsection{A Physical understanding of Eq. (\ref{equation:u1periodicity})} \label{sub1}

Let us consider the following two tori $\mathbb{T}_{0x}$ and $\mathbb{T}_{yz}$. The former is formed by imaginary time direction and $x$-direction. The latter is formed by $y$- and $z$- directions. In both tori, we have the following constraints due to  the condition (\ref{equation:higgsed}):
 \begin{align}
\int \!\!\!\!\! \int_{\mathbb{T}_{0x}} b_{0x} d\tau dx=2\pi\times \mathcal{N}_{0x}\,, \int \!\!\!\!\!\int_{\mathbb{T}_{yz}} b_{yz} dy dz=2\pi\times \mathcal{N}_{yz}\,,\nonumber
\end{align}
where $\tau$ is imaginary time.
Note that, above two integrals are performed at fixed $y,z$ and fixed $\tau,x$ respectively. Since the integrals in both L.H.S. are smooth functions of space-time, the integers $\mathcal{N}_{0x}$ ($\mathcal{N}_{yz}$) should be independent on the coordinates $y$ and $z$ ($\tau$ and $x$).

 Then, one may reformulate $b\wedge b$ term as:
\begin{align}
&i\frac{\Lambda}{16\pi}\int d^4x\ep b_{\mu\nu}b_{\lambda\rho}\nonumber\\
=&i\frac{\Lambda}{16\pi}\int \!\!\!\!\!\int_{\mathbb{T}_{0x}} d\tau dxb_{0x}\int \!\!\!\!\!\int_{\mathbb{T}_{yz}} dydz b_{yz}\epsilon^{0xyz}+ (x0yz)+ (0xzy)\nonumber\\
&+ (x0zy)+(yz0x)+(yzx0)+(zy0x)+(zyx0)\nonumber\\
=&i\frac{8\Lambda}{16\pi}\int \!\!\!\!\!\int_{\mathbb{T}_{0x}} d\tau dxb_{0x}\int \!\!\!\!\!\int_{\mathbb{T}_{yz}} dydz b_{yz}=i \frac{\Lambda}{2\pi} {2\pi}\mathcal{N}_{0x}{2\pi}  {\mathcal{N}}_{yz} \nonumber\\
=&i{2\pi\Lambda}\mathcal{N}_{0x}   {\mathcal{N}}_{yz} \,.\label{eq:deriveshift}
 \end{align}
 By noting that the partition function is invariant if $\int \mathcal{L}$ is shifted by $2\pi$ in a quantized  theory,  we end up with the   periodicity shift shown in Eq. (\ref{equation:u1periodicity}).

  \subsection{A derivation of Eq. (\ref{equation:quantizationzn})}\label{sub2}
  The condition (\ref{equation:higgsed}) leads to $\mathcal{N}_{0x}$ and $\mathcal{N}_{yz}$ (see Appendix \ref{sub1}) quantized at $1/N$ such that the periodicity of $\Lambda$ is:
\begin{align}
 \Lambda\rightarrow \Lambda+N^2\,.\label{equation:periodicityzn}
 \end{align}

On the other hand, as a discrete gauge theory, $B\wedge B$ term should also be invariant up to $2\pi$ under large gauge transformation: $B_{\mu\nu}\rightarrow B_{\mu\nu}+\delta B_{\mu\nu}\,,
$ where $\delta B_{\mu\nu}$ satisfies: $ \int_{\mathcal{S}} \delta B_{\mu\nu} \mathcal{S}^{\mu\nu}=2\pi \times \text{integer}\,.
 $ (there is no implicit summation over indices $\mu,\nu$ here.)
Without loss of generality, let  us still consider the pair of tori $\mathbb{T}_{0x}$ and $\mathbb{T}_{yz}$:
$ \int_{\mathbb{T}_{0x}} \delta B_{0x} d\tau dx=2\pi\times \widetilde{\mathcal{N}}_{0x}\,, \int_{\mathbb{T}_{yz}} \delta B_{yz} dy dz=2\pi\times \widetilde{\mathcal{N}}_{yz}\,,$
The additional terms $\delta S$ arising from the large gauge transformation are collected as follows:
\begin{align}
&S+ \delta S=\nonumber\\
&i\frac{\Lambda}{16\pi}\int \!\!\!\!\!\int_{\mathbb{T}_{0x}} d\tau dx(B_{0x}+\delta B_{0x})\int \!\!\!\!\!\int_{\mathbb{T}_{yz}} dydz (B_{yz}+\delta B_{yz})\epsilon^{0xyz}\nonumber\\
&+ (x0yz)+ (0xzy) + (x0zy)+(yz0x)+(yzx0)\nonumber\\&~~~~~+(zy0x)+(zyx0)\nonumber\\
=&S+\frac{8\Lambda}{16\pi} (\frac{2\pi}{N}\mathcal{N}_{0x}2\pi\widetilde{\mathcal{N}}_{yz}+2\pi\widetilde{\mathcal{N}}_{0x}\frac{2\pi}{N} {\mathcal{N}}_{yz}+2\pi\widetilde{\mathcal{N}}_{0x}2\pi\widetilde{\mathcal{N}}_{yz})  \nonumber\\
=&S+2\pi \Lambda(\frac{1}{N}\mathcal{N}_{0x} \widetilde{\mathcal{N}}_{yz}+\frac{1}{N}\widetilde{\mathcal{N}}_{0x}{\mathcal{N}}_{yz}+\widetilde{\mathcal{N}}_{0x}\widetilde{\mathcal{N}}_{yz})  \,. \label{equation:largegauge}
\end{align}
To keep the quantum theory invariant under the large gauge transformation, $\delta S$ must equal to integer$\times 2\pi$, leading to the   quantization condition (\ref{equation:quantizationzn}).


\begin{thebibliography}{99}
\bibitem{explainspt} In this paper, the shorthand ``SPT'', when it appears alone, always denotes a state in boson/spin systems unless otherwise specified, such as ``fermionic SPT''.
\bibitem{footnotewencoho} It should be pointed out that the relation between these different mathematical tools (i.e. group cohomology and cobordism) has been studied in  \cite{Wencoho} and those states that were ``beyond group cohomology classification'' now are still classified by a more generalized cohomology theory.
\bibitem{footnoteroots} By $\mathbb{Z}_2\times \mathbb{Z}_2$, we mean that a pair of $\mathbb{Z}_2$ indices can be defined and leads to $2\times2=4$ distinct phases labeled by $(0,0)$\,,$(1,0)$\,,$(0,1)$\,,$(1,1)$. Among them, $(0,0)$ is trivial SPT state, and, $(1,0)$ and $(0,1)$ are two distinct root states. The last one, i.e. $(1,1)$ can be easily achieved by simply putting the two root states together in 3d space. This is the intuitive reason why $(1,0)$ and $(0,1)$ are called ``root states''.
\bibitem{remarkcm} The non-compactness of $a$ and $b$ can be understood in the discussions below Eq. (55) in Ref. \onlinecite{3d3} where $a$ and $b$ are integrated out in a usual Gaussian way which implies that singularities of $a$ and $b$ are   not considered such that both are non-compact.
\bibitem{wedge_convention}Here, for readers who are unfamiliar with the basic conventions `$\wedge$', we list some of them here. $\text{d}a\equiv f$
which is a 2-form field strength of 1-form gauge field $a$. Thus, $\int b\wedge \text{d}a=\int b\wedge f=\int d^4x\frac{b_{\mu\nu}}{2}\frac{f_{\lambda\rho}}{2}\ep=\int d^4x\frac{1}{2}b_{\mu\nu} \partial_\lambda a_\rho\ep$; $\int b\wedge b=\int d^4x \frac{b_{\mu\nu} }{2}\frac{b_{\lambda\rho}}{2}\ep=\int d^4x \frac{1}{4}b_{\mu\nu} {b_{\lambda\rho}}\ep$; $\int a\wedge \text{d}a=\int a\wedge f=\int d^3x a_\mu\frac{ f_{\nu\lambda}}{2}\epsilon^{\mu\nu\lambda}=\int d^3x a_\mu \partial_\nu a_\lambda \epsilon^{\mu\nu\lambda}$.
\bibitem{footVS} Note that, in \cite{3d3}, the coefficient   of the topological BF term is incorrect for SPT states which by definition have unique ground state on a three-torus.
\end{thebibliography}
\end{document}